\documentclass[aps,amsmath,amssymb,prd,showpacs,floatfix,preprint,superscriptaddress,nofootinbib,12pt]{article}
\usepackage{jheppub}
\usepackage{bm}
\usepackage{ulem}

\usepackage{axodraw4j}
\usepackage{pstricks}
\usepackage{color}

\DeclareMathOperator{\arctanh}{arctanh}

\newcommand{\p}{\partial}
\newcommand{\Tr}{\text{Tr}}

\newcommand{\Li}{\text{Li}}

\usepackage{epstopdf}

\allowdisplaybreaks[3]

\begin{document}

\preprint{EFI-15-33}

\title{Dense Chern-Simons Matter with Fermions at Large $N$}

\author[]{Michael Geracie,}
\author[]{Mikhail Goykhman}
\author[]{and Dam T. Son}
\affiliation[]{Kadanoff Center for Theoretical Physics, Enrico Fermi Institute and Department of Physics The University of Chicago, Chicago, IL 60637 USA}
\emailAdd{mgeracie@uchicago.edu, goykhman@uchicago.edu, dtson@uchicago.edu}

\abstract{
In this paper we investigate properties of Chern-Simons theory coupled to massive fermions in the large $N$ limit. We demonstrate that at low temperatures the system is in a Fermi liquid state whose features can be systematically compared to the standard phenomenological theory of Landau Fermi liquids.
This includes matching microscopically derived Landau parameters with
thermodynamic predictions of Landau Fermi liquid theory.
We also calculate the exact conductivity and viscosity tensors at zero temperature and finite chemical potential. In particular we point
out that the Hall conductivity of an interacting system is not entirely accounted
for by the Berry flux through the Fermi sphere.
Furthermore, investigation of the thermodynamics in the non-relativistic limit reveals novel phenomena at strong coupling. As the 't Hooft coupling $\lambda$ approaches 1, the system exhibits an extended intermediate temperature regime in which the thermodynamics is described by neither the quantum Fermi liquid theory nor the classical ideal gas law.  Instead, it can be interpreted as a weakly coupled quantum Bose gas.}

\maketitle

\section{Introduction}\label{sec:intro}

The study of large $N$ relativistic Chern-Simons theories at rank $k$ has attracted a great deal of attention in recent years. These theories have proven to be an exceptionally rich, exhibiting a great deal of non-trivial physics, yet still permit exact computations of many relevant quantities at arbitrarily strong coupling. This was initially demonstrated in \cite{Giombi:2011kc}, which performed an exact analysis of the self-energy and spectrum of higher spin currents in the pure fermionic theory. Later, results were generalized to finite chemical potential in \cite{Yokoyama:2012fa}\footnote{The finite temperature calculations in both these works do not account for holonomies. The necessary corrections may be found in \cite{Aharony:2012ns}.}. Later, the exact equation of state was computed as a function of $T$ and $\mu$ in both the bosonic and fermionic cases \cite{Aharony:2012ns}. A number of multi-point functions of currents have also borne exact analysis using these techniques \cite{Aharony:2012nh,GurAri:2012is}. Recently, the 2 to 2 $S$-matrices have been solved for in the purely bosonic, purely fermionic, and $\mathcal N = 1$ supersymmetric cases \cite{Jain:2014nza,Inbasekar:2015tsa} assuming the theory satisfies an interesting modification of naive crossing symmetry required by the presence of a Chern-Simons gauge field.

The models in question are conjectured to be subject to a number of interesting dualities. In the conformal case at zero 't Hooft coupling $\lambda = N / k$, one is merely left with the singlet sector of the critical $U(N)$ vector model, dual to Vasiliev gravity \cite{Klebanov:2002ja}. It has been argued that this duality survives at finite $\lambda$ to a parity violating interpolation between the A-type and B-type Vasiliev theories \cite{Giombi:2011kc,Aharony:2011jz}.

Furthermore, these are theories of anyons that permit both bosonic and fermionic descriptions.
One may pass from one description to the other via a strong-weak coupling duality~%
\cite{Giombi:2011kc,Aharony:2012ns,Aharony:2012nh}. In its most general form, this map was conjectured for arbitrary renormalizable Chern-Simons theories with a fundamental boson and fermion in \cite{Jain:2013gza}. Tests have been undertaken for correlators in \cite{Aharony:2012nh,GurAri:2012is} and free energies in \cite{Aharony:2012ns,Jain:2013py,Takimi:2013zca}. The recently computed $S$-matrices of \cite{Jain:2014nza,Inbasekar:2015tsa} also obey this duality. Large $N$ matter Chern-Simons theories then provide an exact but poorly understood example of bosonization in $2+1$ dimensions. 

The exact solvability of large $N$ Chern-Simons theory makes it an
ideal playground for the investigation of strongly interacting phases
of matter, especially the compressible phases \cite{Sachdev:2012dq}.
In this work we focus on the low-temperature ($\mu / T \gg 1$) phase of the massive fermionic theory at finite chemical potential. Standard fermionic systems typically condense to one of two phases in this limit, a BCS condensate or a Landau Fermi liquid (LFL) \cite{Shankar:1993pf}  (though more exotic states are known to occur in holographic theories \cite{Lee:2008xf,Cubrovic:2009ye}).
In our case, the second option seems a likely candidate. A calculation of the entropy density at low $T$ (equation (\ref{LowTempcV})) yields
\begin{align}
	s =  \frac{\pi}{6} N (1 - \lambda^2 )\mu T + O ( T^2) ,
\end{align}
that of a quantum liquid of interacting fermions.
The minimally coupled large $N$ fermionic theory then offers the tantalizing possibility of realizing a Fermi liquid state whose properties may be exactly computed from microscopics, even at strong coupling.
(A notable example of a Fermi liquid with strong coupling is liquid helium-3,
where the Landau parameters $F_0$ and $F_1$ can be numerically much larger
than one \cite{Helium3book}.)

Motivated by this prospect
we demonstrate via an exact computation in the large $N$ limit that the low-temperature state of fermions minimally coupled to a Chern-Simons gauge field is a stable Landau
Fermi liquid for all values of the coupling constant $\lambda$.
We begin with a brief review of Landau Fermi liquid theory in section \ref{sec:LFLReview}. In section \ref{sec:Thermo} we introduce large $N$ Chern-Simons theory with fermions and then turn to thermodynamics. We investigate the low temperature limit of the known equation of state for later comparison with the results of Landau Fermi liquid theory. To perform this matching, one must take into account the effect of holonomies, which we find suppress the low temperature heat capacity of this model in comparison with a standard Fermi-liquid. In section \ref{sec:DegeneracyTemp} we consider the quantum and classical limits of this system and demonstrate the existence of a novel intermediate regime that only exists at strong coupling. This region appears to exibit a linear specific heat with a slope differing from the predictions of LFL theory. As $\lambda$ approaches 1 this behavior is valid over an ever wider range and the ideal gas regime is pushed to infintely high temperatures at fixed particle-number density.
In section \ref{sec:LandauParameters}, we provide an exact computation of the Landau parameters and demonstrate consistency with the thermodynamic results. 

We also take the opportunity to furnish examples of the behavior of linear response coefficients that are usually not accessible at strong coupling. In section \ref{sec:Conductivity} we perform an exact computation of the conductivity tensor. The result for the Hall conductivity in particular indicates the need to augment LFL theory to properly capture parity odd transport, a problem that has only started to be addressed \cite{Jingyuan2015}.

Finally, in section \ref{sec:Viscosity} we undertake a linear response analysis of the viscosities, obtaining the bulk, shear and Hall viscosities at leading order in $N$. We take the non-relativistic limit of the Hall viscosity and find that it takes the form of a gas of anyons with statistics induced by the Chern-simons interaction
\begin{align}
	\eta_H = \mp \frac{\hbar}{4} (1 \mp \lambda ) n .
\end{align}
Here the sign denotes the relative sign of the 't Hooft coupling and the fermion mass.
Note that $\pm \frac\hbar2(1 \mp \lambda)$ is the angular momentum of an anyon with
statistics $\theta=\pm \pi(1\mp\lambda)$ \cite{Wilczek:1981du}, this formula is
consistent with the linear relationship between the Hall viscosity and the
angular momentum density, which 
has been previously demonstrated by adiabatic methods and confirmed in various non-relativistic gapped states of matter \cite{Read2009,Read2011}.
It also holds for the chiral superfluids \cite{Hoyos:2013eha}.
Whether the relationship continues to hold for general ungapped systems
requires further study.

\section{Review of Landau Fermi Liquid Theory}\label{sec:LFLReview}

In this section we review the fundamentals of Landau Fermi liquid theory to set the stage and establish notation. LFL theory is a phenomenological theory of finite density fermions at ultra-low temperatures, motivated by the intuition that the only relevant exciations in this regime are those of the
quasiparticles in the immediate vicinity of the Fermi surface. 
For a systematic introduction to the subject
we refer the reader to \cite{abrikosov1975methods,baym2008landau,pitaevskii1980statistical,negele1988quantum}\footnote{We will focus on the relativistic, $N$ species, $2+1$ dimensional case relevant to us and so our formulas will occasionally differ from those found in these texts. In all cases however they may be obtained by the techniques explained therein.}. Relativistic aspects of Landau Fermi liquid theory were considered in \cite{baym1976landau}.

Traditionally, the theory is characterized by the so-called Landau parameters. These parameters in hand, one can calculate various thermodynamic observables and long-wavelength modes\footnote{As expected, since these identities are given rigorous microscopic justification on general grounds in \cite{abrikosov1975methods,baym2008landau,pitaevskii1980statistical,negele1988quantum}. However it is worth to point out that it has been argued that a single additional parameter, the Berry flux through the Fermi surface, is required to account for Hall conductivity \cite{Haldane:2004zz}.
We study partity-odd transport in section \ref{sec:Conductivity}.}. We study the thermodynamics of
the large-$N$ Chern-Simons theory with fermions in section \ref{sec:Thermo}.
We will see that the thermodynamic results of section \ref{sec:Thermo} and the
microscopically derived Landau parameters found in section \ref{sec:LandauParameters} are consistent once the effect of holonomies on thermodynamics is taken into account.


\subsection{Quasiparticle Interaction and the Landau Parameters}
Consider a system of interacting fermions in $2+1$ dimensions at finite
chemical potential $\mu$ and temperature $T$ in the low-temperature limit $T/\mu\ll  1$.
LFL theory assumes this system may be described in terms
of interacting quasiparticles
labeled by their momentum $\mathbf p$ where $|\mathbf p| \sim p_F$.\footnote{In 3+1 dimensions quasiparticles are also labeled by their spin.}
The quasiparticle occupation number in momentum space $n ( \mathbf p )$ is a perturbation about the zero temperature distribution $n_0 (\mathbf p)$ with Fermi surface at $|\mathbf p| = p_F$,
\begin{align}
	n(\mathbf p ) = n_0 (\mathbf p ) + \delta n (\mathbf p ) .
\end{align}
The low-lying spectrum of excitations consists of individual quasiparticles and quasiholes, as well as possible collective modes involving long wavelength deformations of the Fermi surface. 

Denote the energy cost of adding a single quasiparticle with momentum ${\bf p}$ by $\varepsilon(\bf p)$.  Due to interactions, $\varepsilon ( \bf p )$ will in general depend on the occupation 
number $n(\bf p)$. To first order we then have
\begin{align}\label{fDef1}
	\varepsilon ( \mathbf p ) = \varepsilon_0 ( \mathbf p ) + \int \frac{d^2 p'}{(2 \pi )^2} f (\mathbf p , \mathbf p' ) \delta n ( \mathbf p')
\end{align}
for some function $f(\mathbf p , \mathbf p')$ that parameterizes the strength of interactions between quasi-particles at different points in momentum space.
We can now identify the Fermi velocity and effective mass of single particle excitations in the vicinity of the Fermi surface
\begin{align}\label{vF}
	v_F=\frac{\p \varepsilon_0 (\bf p )}{\p |{\bf p}|}\Bigg|_{|{\bf p}|=p_F},
	&& m^\star=\frac{p_F}{v_F}\,.
\end{align}

When multiple species of fermions are present, the quasiparticle energy and occupation number are Hermitian operators in the internal space and (\ref{fDef1}) becomes
\begin{align}\label{fDef}
	\varepsilon^i{}_j ( \mathbf p ) = \varepsilon_0^i{}_j ( \mathbf p ) + \int \frac{d^2 p'}{(2 \pi )^2} f^i{}_j,{}^l{}_k (\mathbf p , \mathbf p' ) \delta n^k{}_l ( \mathbf p') \,.
\end{align}
For us, there is only a single on-shell spin degree of freedom to keep track of and the internal space is simply color space.\footnote{We hope that context will be sufficient to distinguish between color indices and spatial coordinate indices, both of which we shall denote with lower case Latin letters.}

For us, there is only a single on-shell spin degree of freedom to keep track of and the internal space is simply color space. We then decompose the interaction strength 
 into direct and exchange channels
\begin{align}
	f^i{}_j,{}^l{}_k (\mathbf p , \mathbf p' ) = f^{(d)} ( \mathbf p , \mathbf p' ) \delta^{i}{}_j \delta^{l}{}_k + f^{(e)} ( \mathbf p , \mathbf p' ) \delta^{i}{}_k \delta^{l}{}_j .
\end{align}
This must be symmetric under the exchange $\mathbf p  \leftrightarrow \mathbf p'$, $i \leftrightarrow l$, $j \leftrightarrow k$ and so
\begin{align}
	f^{(d)} (\mathbf p , \mathbf p') = f^{(d)} ( \mathbf p' , \mathbf p ) ,
	&&f^{(e)} (\mathbf p , \mathbf p') = f^{(e)} ( \mathbf p' , \mathbf p ) .
\end{align}
We will assume that the $U(N)$ symmetry that acts on the ``color" index $i$ is unbroken. In this case $\varepsilon_0^i{}_j$ is proportional to the identity matrix and the definitions (\ref{vF}) are applicable. In what follows we will often suppress color indices when their placement is obvious.

In the low-temperature limit, we expect excitations to be localized near the Fermi surface, so that we are free to take $|{\bf p}|=|{\bf p'}|=p_F$.
The quasi-particle
interaction function $f$ then depends only on the angle $\theta$
between ${\bf p}$ and ${\bf p'}$. It is convenient to parameterize this function by introducing the Landau parameters
$F^{(d)}_n, F^{(e)}_n$
\begin{align}
	\label{fExpansion}
	&f^{(d)} ( \theta ) = \frac{2 \pi}{N m^\star} \left( F^{(d)}_0 + 2 \sum_{n=0}^\infty  F^{(d)}_n \, \cos n \theta  \right) , \nonumber \\
	&f^{(e)}(\theta)= \frac{2 \pi}{N m^\star} \left( F^{(e)}_0 + 2 \sum_{n=0}^\infty  F^{(e)}_n \, \cos n \theta  \right).
\end{align}
Normalization by the density of states at the Fermi surface
\begin{equation}
\nu (\epsilon_F ) = \int_{|{\bf p}|=p_F}\frac{d^2p}{(2\pi)^2}= \frac{N m^\star}{2 \pi}\,
\end{equation}
is conventional.

\subsection{Common Observables in Landau Fermi Liquid Theory}
In this section, we briefly touch on several observables that may be computed within LFL theory.
These observables will be the basis for comparison with field theoretic results found in section \ref{sec:LandauCalc}.
Consider first the total energy flux carried by some near-equilibrium distribution of quasi-particles
\begin{align}
	\frac V 2 \int \frac{d^2 p}{(2 \pi)^2} \Tr \left( \left( \varepsilon ( \mathbf p )\frac{\partial \varepsilon (\mathbf p)}{\partial p_i} +\frac{\partial \varepsilon (\mathbf p)}{\partial p_i}  \varepsilon ( \mathbf p ) \right) n ( \mathbf p) \right).
\end{align} 
By Lorentz invariance, the energy is merely the time component of an energy-momentum four-vector, so the energy flux must be equal to the total momentum flux
\begin{align}
	V \int \frac{d^2 p}{(2 \pi )^2} p^i \Tr \left(n ( \mathbf p ) \right) .
\end{align}
Expanding to first order in $\delta n ( \mathbf p)$ then yields a simple equation determining the effective mass induced by interactions
\begin{align}\label{effective mass}
	m^\star =\mu \left(1+ F^{(d)}_1 + \frac 1 N F^{(e)}_1 \right)\,,
\end{align}
where we have evaluated at the Fermi surface and used $\mu = \varepsilon_F$.

We now turn to the isothermal inverse compressibility $\kappa^{-1} = n^2 \left( \frac{\partial \mu}{\partial n} \right)_T$. This is determined by both the zeroth and first Landau parameters. To obtain $\left( \frac{\partial \mu}{\partial n} \right)_T$, one introduces a small perturbation in the total density
$\delta n = \int \frac{d^2 p}{(2 \pi)^2} \delta n ( \mathbf p )$. We can then calculate the shift in the chemical potential arising from the shift in the location of the Fermi surface and from interactions between the perturbation and the Fermi sea. Using (\ref{effective mass}) one finds the isothermal inverse compressibility to be given by
\begin{align}\label{compressibility}
	\kappa ^{-1} = \frac{2 \pi n^2}{N m^\star} \left(1+ F^{(d)}_0 + \frac 1 N F^{(e)}_0 \right)\,.
\end{align}

Finally, it was pointed out in \cite{Haldane:2004zz} that the Landau parameters are not sufficient to describe parity odd-transport in $2+1$ dimensions. In this case, the Hall conductivity picks up an essential contribution from the Berry flux through the Fermi ball
\begin{align}\label{HallConductivity}
	\sigma_H = \int \frac{d^2 p}{(2 \pi)^2} \Tr ( \mathcal  F( \mathbf p ) n(  \mathbf p ) ) = \frac{1}{4\pi^2} \oint_{p_F} \Tr ~ \mathcal A ,
\end{align}
where $\mathcal A = - i \psi^\dagger d \psi$ is the Berry connection and $\mathcal F = d \mathcal A$ is the associated flux density.

\subsection{Microscopics}

The Landau parameters may be calculated directly within quantum field theory \cite{baym1976landau}, the fundamental object of interest being the on-shell four-point function  
\begin{align}\label{V def}
	V^i{}_l,^k{}_j(p,k,q) = \frac{1}{(u^\dagger u)^2} u_\alpha ( p + q)  u_\gamma (k ) V^\alpha{}_\delta,{}^\gamma{}_\beta;^i{}_l,^k{}_j(p,k,q) \bar u^\beta ( k + q ) \bar u^\delta (p) .
\end{align}
Here $u_\alpha (p)$ are the on-shell quasiparticle spinors.
\begin{align}
	V^\alpha{}_\delta,{}^\gamma{}_\beta;^i{}_l,^k{}_j(p,k,q) = \left \langle  \bar\psi^{\alpha i} (- p - q) \bar\psi^{\gamma k} (- k) \psi_{\delta l} (p)  \psi_{\beta j} ( k + q ) \right \rangle_\text{1PI}
\end{align}
is the 1PI four point function, represented by the diagram\footnote{Our conventions for spinors are as follows.
On Feynman diagrams an outgoing line corresponds to spinor $\psi_{\alpha i}$ with lower indices,
and incoming line corresponds to Dirac-conjugate spinor $\bar \psi^{\alpha i}$ with upper indices.}.

\begin{center}
\fcolorbox{white}{white}{
   \scalebox{.5}{
  \begin{picture}
  (456,184) (19,-91)
    \SetWidth{1.0}
    \SetColor{Black}
    \GOval(224,8)(80,40)(0){0.882}
    \Line[arrow,arrowpos=0.5,arrowlength=8,arrowwidth=3,arrowinset=0.2](256,56)(408,56)
    \Line[arrow,arrowpos=0.5,arrowlength=8,arrowwidth=3,arrowinset=0.2](408,-42)(256,-42)
    \Line[arrow,arrowpos=0.5,arrowlength=8,arrowwidth=3,arrowinset=0.2](48,56)(192,56)
    \Line[arrow,arrowpos=0.5,arrowlength=8,arrowwidth=3,arrowinset=0.2](192,-42)(48,-42)
    \Text(112,72)[lb]{\Large{\Black{$p+q$}}}
    \Text(320,72)[lb]{\Large{\Black{$k+q$}}}
    \Text(112,-75)[lb]{\Large{\Black{$p$}}}
    \Text(328,-75)[lb]{\Large{\Black{$k$}}}
    \Text(16,56)[lb]{\Large{\Black{$\alpha\;\; i$}}}
    \Text(424,56)[lb]{\Large{\Black{$\beta\;\; j$}}}
    \Text(16,-42)[lb]{\Large{\Black{$\delta\;\; l$}}}
    \Text(424,-42)[lb]{\Large{\Black{$\gamma\;\; k$}}}
    \Text(211,7)[lb]{\Large{\Black{{\bf 1 PI}}}}
  \end{picture}
  }
}
\end{center}

The interaction function is then obtained by taking the scattered particles to lie on the Fermi surface $|\mathbf p| = | \mathbf k | = p_F$ and the exchange momentum $q$ to zero in the ``rapid" limit
\begin{align}\label{InteractionFunction}
	f^i{}_j;^k{}_l ( \theta ) = Z^2 \lim_{q^0 \rightarrow 0} \lim_{\mathbf q \rightarrow \mathbf 0} V^i{}_j;^k{}_l ( p , k , q) .
\end{align}
Here $\theta$ is the angle between the scattered quasiparticles and $Z$ the wavefunction renormalization. The latter is defined by the quasiparticle propagator
\begin{equation}
	\label{LFLFermionicPropagator}
	G(p)^{\; \alpha}_{\;\;\; \beta}\,=\,
	\frac{Z}{\omega- v_F (|{\bf p}|- p_F)+i\,\epsilon\, {\rm sgn}\, (|{\bf p}|-p_F)}
	\frac{\bar u^\alpha \, u_\beta}{u^\dagger u} \bigg|_{|{\bf p}|=p_F}\,,
\end{equation}
expanded in the vicinity of the Fermi surface, $\omega=0$, $|{\bf p}|=p_F$. Expanding the propagator (\ref{prop}) and using the exact wavefunctions (\ref{UonFermiSurface}) we find that in large $N$ CS theory with fundamental fermions, the quasiparticle propagator takes precisely this form with $Z=1$.

\section{Thermodynamics of Large $N$ Chern-Simons Systems with Fermions}\label{sec:Thermo}

We now turn to the theory which interests us in this paper, namely, large $N$ Chern-Simons theory coupled to massive fundamental fermions. As reviewed in the introduction, this is a remarkable model insofar as it exhibits a great deal of non-trivial physics that may be extracted exactly at arbitrary values of the  coupling. We introduce this theory in section \ref{sec:CSintro} and review several key results, including the known exact equation of state at finite temperature and density. In section \ref{sec:LowTemp} we set the stage for our later examination of the LFL state by considering the low temperature limit.

The Fermi liquid state is a strongly quantum regime where the specific heat is linear in the temperature $c_v \sim T$ and the slope is determined by the effective mass. 
It was demonstrated in \cite{Aharony:2012ns} that holonomies about the thermal circle need to be accounted for to correctly capture the thermodynamics. We observe in section \ref{sec:Holonomies} that at low temperatures the effect of holonomies is to dampen the specific heat of
the system.

In section \ref{sec:DegeneracyTemp} we demonstrate that the transition temperature above which the gas effectively becomes classical diverges as the coupling is increased, while the quantum degeneracy temperature remains fixed. At strong coupling then a new extended region emerges which is neither quantum nor classical in nature. We observe this regime numerically and find that it is also characterized by a linear specific heat, but with a slope differing from that of the Large $N$ CS Fermi-liquid.

\subsection{A Review of Large $N$ Chern-Simons Theory with Fundamental Fermions}\label{sec:CSintro}

The theory we will concern ourselves with for the duration of this paper is that of $N$ species of fermions in the fundamental of $SU(N)$, coupled to Chern-Simons theory at level $k$ and rank $N$. In the 't Hooft limit
\begin{align}
	k,N \rightarrow \infty\, ,
	&& \lambda = \frac{N}{k} \text{~~fixed} ,
\end{align}
this theory becomes exactly solvable. 

The Lagrangian density at finite chemical potential is given by
\begin{equation}
{\cal L}=N \left( \frac{i}{4\pi \lambda}\,\epsilon^{\mu\nu\lambda}\,{\rm Tr}\,\left(A_
\mu\p_\nu A_\lambda-\frac{2i}{3}A_\mu A_\nu A_\lambda\right)+\bar\psi \gamma^\mu D_\mu \psi
+m\,\bar\psi\psi\,-\mu\bar\psi \gamma^3 \psi\right).\label{FullLagrangian}
\end{equation}
Here we work in Euclidean space, $\bar \psi^\alpha = (\psi_\alpha)^*$, $D_\mu \psi=(\p_\mu -iA_\mu)\psi$, $\mu$ is the chemical potential and $m$ the bare mass of the fermion. The coupling constant $\lambda$ is a continuous parameter in the 't Hooft limit and can take any value $|\lambda | < 1$. Near $| \lambda | = 1$ the theory approaches infinite coupling.
We shall find it convenient to fix $\lambda$ and $\mu$ to be positive by application of $C$ and $P$. The mass may then have any sign. When results depend on the sign of the mass, we will indicate the positive mass result by the upper sign and the negative mass result with the lower sign\footnote{If one does not fix the sign of $\lambda$ this difference is in the relative sign of $m_0$ and $\lambda$.}. 

Following \cite{Giombi:2011kc}
we shall represent the gamma matrices as
\begin{align}
	\gamma^\mu = 
	\begin{pmatrix}
		\sigma^1 & \sigma^2 & \sigma^3
	\end{pmatrix},
\end{align}
work in light-cone coordinates $x^\pm = \frac{1}{ \sqrt{2}} (x^1 \pm i x^2)$, and
set light-cone gauge $A_-=0$. In this gauge, the non-vanishing components
of the gauge field propagator $G_{\mu \nu} (p)$ are given by
\begin{equation}
	G_{+3}(p)=-G_{3+}(p)= \frac{4\pi i \lambda}{p_-}\,,
	\label{GaugePropagator}
\end{equation}
which is exact in the large $N$ limit. Here and in what follows we will often suppress factors of $N$ when their placement is obvious.

The full fermionic propagator is 
\begin{align}
	G(p) &= \frac{1}{i\tilde p_\mu \gamma^\mu+\Sigma(p)} ,
	\label{FermionPropagator}
\end{align}
where we have denoted $\tilde p_\mu=p_\mu+i\mu\delta_{\mu\,3}$ and $G(p)^\alpha{}_\beta = \left \langle \psi_\beta (p)  \bar \psi^\alpha (- p) \right \rangle$.
In the large $N$ limit, the self-energy $\Sigma(p)$ satisfies a recursion relation which has been solved at finite temperature and chemical potential \cite{Aharony:2012ns,Jain:2013py,Takimi:2013zca,Jain:2013gza}.
One finds that in the light-cone gauge, $\Sigma (p)$ is of the form
\begin{align}
	\label{SelfEnergyMatrix}
	\Sigma (p) &=\Sigma_I (p) I+\Sigma_+ (p) \gamma^+ ,
	&&\text{where} &&
	\Sigma_+(p) =\frac{ c^2_0 - \Sigma^2_I (p) }{ p^2_s} ip_+ .
\end{align}
Given this, one may rewrite the propagator as
\begin{align}\label{prop}
	G(p) = \frac{\Sigma_I - i \gamma^+ ( p + \Sigma )_+ - i \gamma^- p_- - i \gamma^3 \tilde p_3 }{\tilde p^2 + c_0^2},
\end{align}
from which we see that $c_0$ is the pole mass, determined \cite{Jain:2013gza} by the gap equations\footnote{Note we have made a choice of sign in (2.12) of \cite{Jain:2013gza}.}, 
\begin{align}\label{gapEquations}
	&\hat c_0 = \hat m + 2 \lambda \mathcal C, \nonumber \\
	&\mathcal C = \frac 1 2 \int d \alpha \rho (\alpha ) \left( \ln 2 \cosh \frac{| \hat c_0 | + \hat \mu + i \alpha}{2} + \ln 2 \cosh \frac{| \hat c_0 | - \hat \mu - i \alpha}{2} \right) .
\end{align}
Any quantity bearing a hat denotes that quantity in units of the temperature, for instance, $\hat \mu = \mu/ T$. The gap equations (\ref{gapEquations}) always have a unique real solution, so that quasi-particles are perfectly stable. This is to be expected in the large $N$ limit, which suppresses internal fermion loops that would lead to decay. We anticipate that finite $N$ effects would introduce a nonzero decay rate.

We use $\alpha$ to denote holonomy eigenvalues about the thermal circle. These have density $\rho (\alpha)$ which approaches the universal form \cite{Aharony:2012ns}
\begin{align}
	\rho ( \alpha ) = 
	\begin{cases}
		\frac{1}{2 \pi \lambda} , & \alpha \in ( - \pi \lambda, \pi \lambda ) , \\
		0 , & \text{otherwise} ,
	\end{cases}
\end{align}
in the thermodynamic limit $\frac{V T^2}{N} \gg 1$, which we shall adopt here.
Finally, having solved for $c_0$, we may find $\Sigma_I (p)$, which is a function of only $p_s$
\begin{align}\label{SigmaI}
	\Sigma_I (p) = m + \lambda  \int d \alpha \rho (\alpha ) \left( \log 2 \cosh \frac{\hat E_p + \hat \mu + i \alpha}{2} + \log 2 \cosh \frac{\hat E_p - \hat \mu - i \alpha}{2} \right)  
\end{align}
where $E_p = \sqrt{p^2_s + c_0^2}$ and $p^2_s = 2 p_+ p_- = p_1^2 + p_2^2$ is the square of the spatial momentum.

The exact free energy is also known at finite temperature and chemical potential and is given by \cite{Aharony:2012ns,Jain:2013gza}
\begin{align}\label{EOS}
	F =\,& \frac{N V T^3}{6 \pi} \bigg( |\hat c_0|^3 - 2 ( | \hat c_0|^2 - \hat m^2) \mathcal C + 2 \lambda \hat m \mathcal C^2   - f_0 \nonumber \\
	&- 3 \int d \alpha \rho ( \alpha ) \int^\infty_{|\hat c_0|} dy y \left( \log (1 + e^{- y - \hat \mu - i \alpha } )+ \log ( 1 + e^{-y + \hat \mu + i \alpha }) \right) \bigg) \,.
\end{align}
Here
\begin{equation}
f_0= | \hat m |^3 \frac{2 \mp \lambda}{(1 \mp \lambda)^2}
\end{equation}
is introduced as a counter-term in the action to subtract off the vacuum energy density.

\subsection{The Low Temperature Limit}\label{sec:LowTemp}
In this section we shall derive the low temperature ($\hat \mu \gg 1$) limit of the expressions found in section \ref{sec:CSintro}. In this limit, up to corrections that are exponentially suppressed in $\hat \mu$, the gap equations (\ref{gapEquations}) reduce to 
\begin{align}
	&\hat c_0 = \hat m + \lambda \max ( | \hat c_0 | , \hat \mu )  .
\end{align}
The self-energy (\ref{SigmaI}) exhibits discontinuous behavior, indicating the presence of a Fermi surface at $p_F = \sqrt{\mu^2 - c_0^2}$ (see also (2.27) of \cite{Yokoyama:2012fa})
\begin{align}\label{SelfEnergy}
	\Sigma_I ( p ) &= m + \lambda \, E_p\, \theta ( E_p - \mu ) + \lambda\, \mu \,\theta ( \mu - E_p )   .
\end{align}

The solution to the gap equation depends on the location of the chemical potential. Below the gap we shall denote $c_0$ as $m_0$ and one finds
\begin{align}\label{m0 mass}
	m_0 = \frac{m}{1 \mp \lambda},
\end{align}
while for $\mu$ greater than $| m_0 |$, one has
\begin{align}\label{c0 mass}
	c_0 = m+ \lambda \mu = m_0 + \lambda \Delta \mu ,
\end{align}
where $\Delta \mu = \mu - | m_0 |$. Recall that the upper sign indicates the result for positive fermion mass.
Equation (\ref{m0 mass}) is then the single particle mass at zero temperature and zero density. When the system is at finite density, self-interactions induce a mass (\ref{c0 mass}).

The Free energy may also be expanded and the holonomy integrals are easily performed in this limit. One finds for $\mu > | m_0 |$, again up to exponentially suppressed corrections,
\begin{align}
\label{FreeEnergy}
	F = \frac{N V}{12 \pi} \bigg( &3 m^2 \mu  + 3 \lambda m \mu^2  + (\lambda^2-1) \mu^3 + \pi^2 ( \lambda^2 -1) \mu T^2 - f_0 + \mathcal O (e^{- \hat \mu})\bigg) .
\end{align}
It is then a simple matter to evaluate the number density, entropy density and specific heat
\begin{align}\label{LowTempcV}
	n = \frac{N p_F^2}{4 \pi}+ \frac{N \pi}{12} (1 - \lambda^2 ) T^2,
	&&s = \frac{N}{6} \pi (1 - \lambda^2) \mu T,
	&&c_V = \frac{N}{6} \pi (1 - \lambda^2) \mu T.
\end{align}
Note that at zero temperature the number density obeys Luttinger's theorem \cite{Luttinger}
\begin{equation}
n =N\,\int_{p_s\leq p_F}\frac{d^2p}{(2\pi)^2}\,.
\end{equation}

From the nonrelativistic limit (\ref{NRp}) we can also easily find the ground state energy of the non-relativistic Chern-Simons Fermi-liquid at finite density.
\begin{align}
	E_{GS}	&= \frac{N V}{4 \pi} | m_0 |    (1 \mp \lambda)  \Delta \mu^2 .
\end{align}

We can then extract the Bertsch parameter \cite{Bertsch}, important in the study of the unitary Fermi gas, and find that for mass the same sign as $\lambda$, it vanishes as the coupling is tuned to infinity
\begin{align}
	E_{GS} &= \xi_B E_{FG}, 
	&&\xi_B = 1 \mp \lambda .
\end{align}
Here $E_{FG}= \frac{NV p_F^4}{16 \pi} = \frac{NV}{4 \pi} | m_0 | \Delta \mu^2 $ is the ground state energy of the free Fermi gas in two spatial dimensions.

The vanishing of the Bertsch parameter as $\lambda\to1$ for positive
mass can be understood easily if one recall that in this limit the
system allows a dual description in terms of weakly coupled bosons.
The bosons are not subject to the Pauli exclusion principle and the
ground state energy is zero when the bosons are not interacting.

\subsection{Holonomies and Statistics}\label{sec:Holonomies}
In this section we discuss an effect of holonomies that will be essential in our verification of the Fermi-liquid state. Namely, that fermions no longer obey a Fermi-Dirac distribution, resulting in a suppression of the specific heat in the quantum regime. To see this, we directly evaluate the mean occupation number from the Green's function
\begin{align}
	n(p_s) &= - \int d \alpha \rho ( \alpha )  \frac{1}{\beta} \sum_{n=- \infty}^{\infty} \Tr \left(  G(p) \gamma^3 \right) \,,
\end{align}
where $\tilde p_3 /T=2 \pi \left( n + 1/2 \right) + i \hat \mu - \alpha$. The sum is over Matsubara
frequencies, shifted by holonomies.
At weak coupling $\lambda \rightarrow 0$, one obtains the standard momentum distribution for relativistic fermions at nonzero temperature and chemical potential \cite{Yokoyama:2012fa}. Restoring the holonomies by the shift $\hat \mu \rightarrow \hat \mu + i \alpha$ one finds
\begin{align}\label{distribution}
	n (p_s) = \frac{N}{2} \int d \alpha \rho ( \alpha ) \left( \tanh \frac{\hat E_p + \hat \mu + i \alpha}{2} - \tanh \frac{\hat E_p - \hat \mu - i \alpha}{2}\right) \,.
 \end{align} 

\begin{figure}
	\centering
	\begin{minipage}{0.45\textwidth}
	\centering
		\includegraphics[width=1\textwidth]{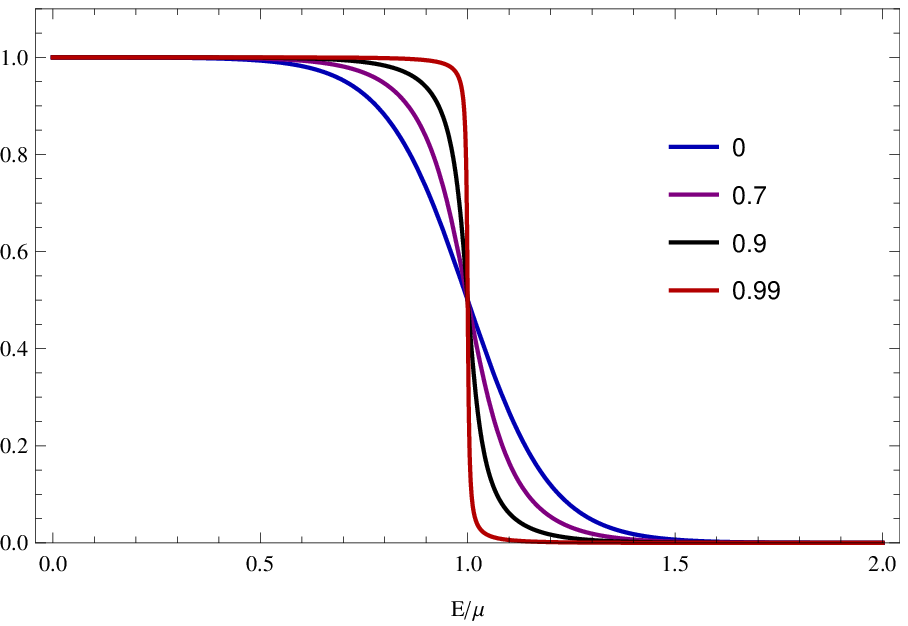}
			\caption{The distribution function (\ref{distribution}) at $T/\mu = .1$ for several different values of $\lambda$. At $\lambda=0$ we have the Fermi-Dirac distribution.}
	\label{fig:FDvsHolonomies1}
	\end{minipage}\hfill
	\begin{minipage}{0.45\textwidth}
	\centering
		\includegraphics[width=1\textwidth]{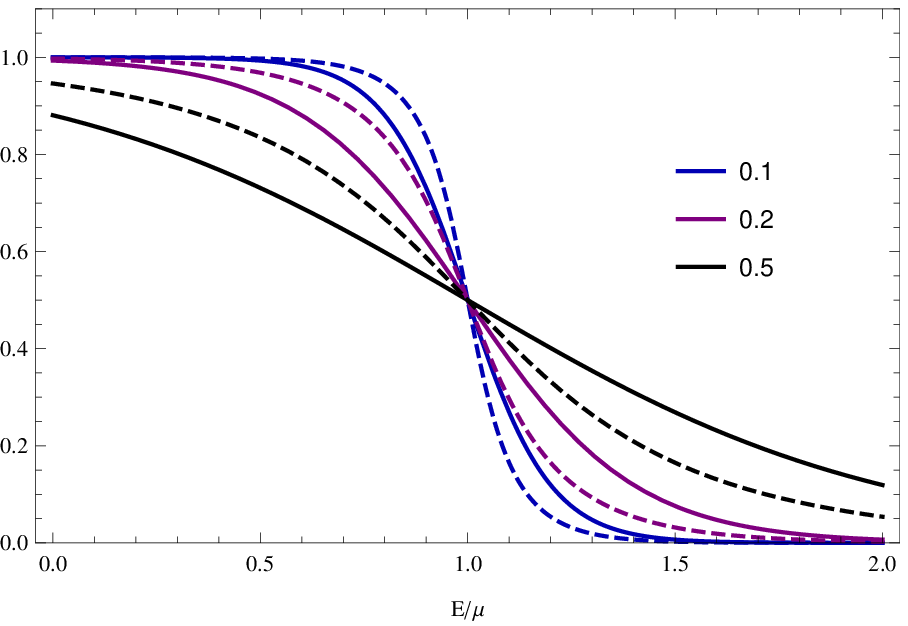}
		\caption{Distribution functions for several values of $T/\mu$ at $\lambda = .7$. The Fermi-Dirac distribution is displayed in solid lines while (\ref{distribution}) is displayed in dashed lines.}
		\label{fig:FDvsHolonomies2}
	\end{minipage}
\end{figure}
 
As seen in figures \ref{fig:FDvsHolonomies1} and \ref{fig:FDvsHolonomies2}, the holonomies make the fermion system ``seem colder'' than a standard Fermi gas at the same temperature, enhancing the tendency of electrons to accumulate below the Fermi momentum.
With fewer fermions excited by heating, we would expect a corresponding suppression in the specific heat of the quantum liquid. Indeed, evaluating the entropy density (see Appendix \ref{app:CvHolonomies} for details)
\begin{align}
	s = - \int \frac{d^2 p}{(2 \pi )^2} \big( n (p_s) \log n (p_s) + ( 1 - n ( p_s) ) \log (1 - n (p_s )) \big),
\end{align}
one finds
\begin{align}\label{CvLowTemp}
	c_v = \frac{N}{6} \pi ( 1 - \lambda^2 ) m^\star T,
\end{align}
compared to $\frac \pi6 N m^\star T$ for a standard Fermi liquid.
We stress that this discrepancy does not invalidate the Landau Fermi liquid expression
for the effective mass $m^\star$ of quasiparticles, but rather is a
consequence of a peculiar quasiparticle distribution function due to the holonomies.

\subsection{A Novel Regime at Strong Coupling}\label{sec:DegeneracyTemp}
In the previous section we discussed some peculiar properties of the large $N$ CS Fermi-liquid state. In this section we investigate the opposing, high temperature limit, in which the gas becomes ideal. As we shall see, as the coupling is tuned to 1, the ideal gas becomes increasingly inaccesible. There is then an extended intermediate regime that exists only at strong coupling for which neither the classical nor quantum descriptions is valid.  

To see this we examine the non-relativistic gas at constant density (throughout this section we refer the reader to appendix \ref{App:Non-Relativistic Thermodynamics} for details). In the non-relativistic limit, the temperature and chemical potential are small compared to the gap energy
\begin{align}
	T, \Delta \mu \ll |c_0| , 
	&&\xi = \frac{\Delta \mu}{T} \qquad \text{arbitrary}.
\end{align}
In this limit the equation of state (\ref{EOS}) becomes
\begin{align}\label{NREOS}
	F = - \frac{N V | m_0 |}{12 \pi} T^2 \left( \pi^2 ( 1 - \lambda^2 ) + 3 \xi^2 \pm 3 \lambda f (\xi ) ( f(\xi) - 2 \xi ) + 6 \int d \alpha \rho ( \alpha ) \Li_2 \left( - e^{- \xi - i \alpha \pm \lambda f ( \xi )}\right) \right),
\end{align}
where $f(\xi)$ determines the pole mass
\begin{align}\label{NRc0}
	c_0 = m_0 + \lambda T f(\xi)
\end{align}
and solves the trancendental equation 
\begin{align}
	f(\xi) = \xi + \frac{1}{\pi \lambda} \text{Im} ~ \Li_2 \left( - e^{- \xi  - i \pi \lambda \pm \lambda f (\xi) }\right) .
\end{align}

Heating this system at fixed particle density $n$, one will eventually enter the classical regime and the specific heat saturates to a constant. To find the range over which this description is valid, we perform a virial expansion of (\ref{NREOS}). We state our results in terms of the pressure $p = - F/V$.
\begin{align}\label{idealGas}
	\frac{p}{nT} = 1 + v_2 \frac{n}{N | m_0 | T} + \mathcal O \left(  \left( \frac{n}{N | m_0 | T}  \right)^2 \right) ,
\end{align}
where $v_2$ is the second virial coefficient
\begin{align}\label{v2}
	v_2&=\frac{\pi \lambda}{\pm 1-\lambda}+ \frac{1}{2} \pi^2 \lambda \cot  \pi \lambda \nonumber \\
	&\rightarrow \mp \frac{\pi}{2} \frac{1}{1-\lambda} 
	\qquad \text{as } \lambda \rightarrow 1.
\end{align}
Note that $v_2$ diverges when $\lambda\to1$.
At strong coupling then, corrections to the ideal gas law $p=nT$ are numerically small only when 
\begin{align}
	T \gg \frac{2 \pi}{1 - \lambda}\frac{n}{N | m_0 |},
\end{align}
which diverges as $\lambda \rightarrow 1$.

\begin{figure}
	\centering
	\begin{minipage}{0.55\textwidth}
	\centering
		\includegraphics[width=1\textwidth]{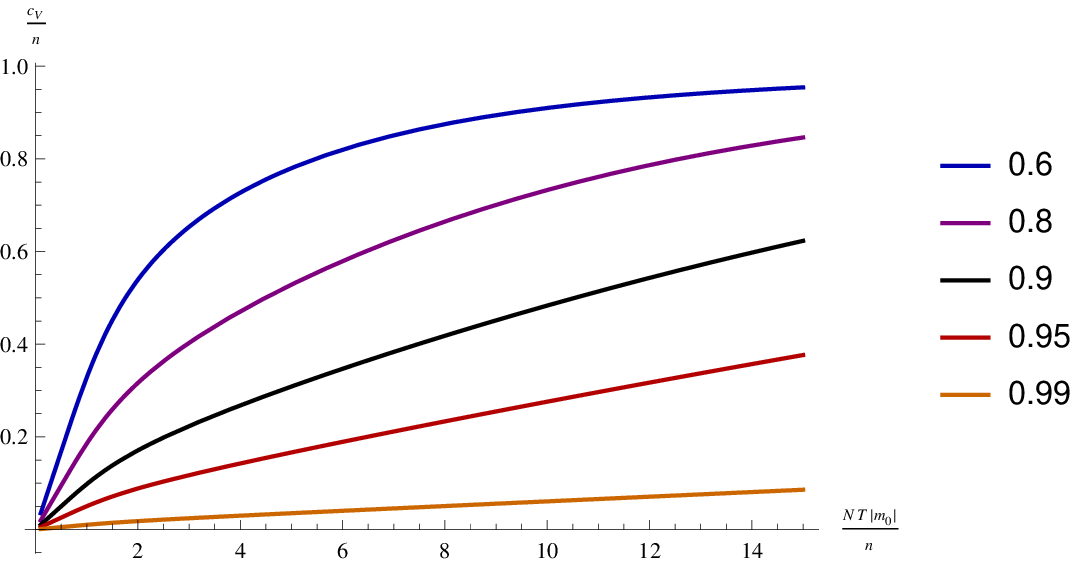}
			\caption{The specific heat for $m_0>0$ at various values of $\lambda$. The transition to the intermediate regime occurs at order $T \approx \frac {n}{N|m_0|}$.}
			\label{fig:SpecificHeat}
	\end{minipage}\hfill
	\begin{minipage}{0.4\textwidth}
	\centering
		\includegraphics[width=1\textwidth]{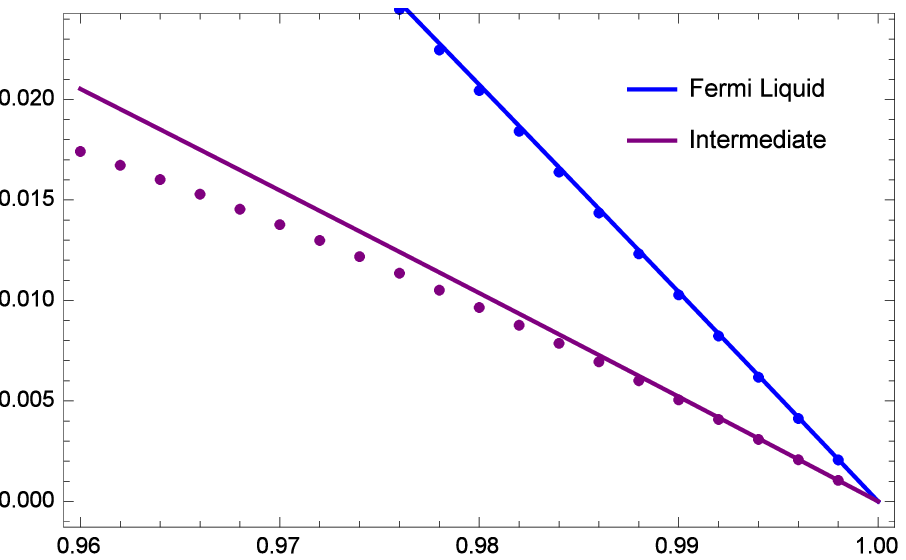}
		\caption{The slopes of $c_V$ in units of $N |m_0|$ of the LFL and intermediate regions as a function of $\lambda$. The linear fits are $\frac \pi 6 (1 - \lambda^2)$ and $\frac \pi 6 ( 1 - \lambda )$ respectively.}
		\label{fig:Slopes}
	\end{minipage}
\end{figure}

On the other hand, a similar expansion in the low temperature limit shows that corrections to the Fermi-liquid specific heat (\ref{CvLowTemp}) are exponentially suppressed and their importance does not depend strongly on the coupling. The system then forms a degenerate liquid when
\begin{align}
	 \pi \frac{n}{N | m_0 | T} \gg 1,
	&&\text{i.e.}
	&& T \ll T_q,
	&&\text{where}
	&& T_q =  \pi \frac{n}{N |m_0 |}.
\end{align}
$T_q$ is simply the degeneracy temperature.
This behavior is denomstrated in the plots above. In figure \ref{fig:SpecificHeat} we plot the specific heat at constant density. The emergence of a second linear regime when $T_q\ll T\ll T_q/(1-\lambda)$ is clear as the coupling is increased. The slopes of the quantum and intermediate regimes are graphed as a function of $\lambda$ in figure \ref{fig:Slopes}. The numerical evidence indicates that the slope of the intermediate regime is half that of the LFL regime as one approaches infinite coupling.


\subsection{Bose-Fermi Duality in Thermodynamics}

What is the nature of the temperature scale $T_q/(1-\lambda)$ and of
the intermediate regime $T_q\ll T\ll T_q/(1-\lambda)$, in which
neither the Fermi liquid nor the classical gas picture work?  A hint
can be obtained from the boson-fermion duality, which maps strongly
coupled fermions into weakly coupled bosons.  Under this duality, the
number of colors of the boson is $N_{\rm boson}=k-N\approx(1-\lambda)
N$, and when fermions are at strong coupling, $1-\lambda\ll1$, the
number of bosonic colors is much smaller than the number of fermionic
colors: $N_{\rm boson}\ll N$.  In the bosonic picture, the number
density per color is much larger than the density per color for
fermions: $n/N_{\rm boson}\gg n/N$, and subsequently the degeneracy
temperature of the boson is also much larger:
\begin{equation}
  T_{\rm bos.deg.} = \frac{n}{N_{\rm boson}m} \approx
    \frac n{(1-\lambda)N m} = \frac{T_q}{1-\lambda}\,.
\end{equation}
Thus the temperature scale $T_q/(1-\lambda)$, mysterious from the
fermionic viewpoint, is simply the degeneracy temperature of the
bosons.  The deviation of $c_v$ from the classical gas value below
$T_q/(1-\lambda)$ is thus the manifestation of the fact that the
bosons behave quantum mechanically below their degeneracy temperature.

This interpretation of the temperature scale $T_q/(1-\lambda)$ is
supported by the virial coefficient at high temperatures.  Using
standard statistical mechanics, one finds that the virial coefficient
of an ideal gas, as defined in Eq.~(\ref{idealGas}), is equal to
$+\pi/2$ for fermions and $-\pi/2$ for bosons.  When $\lambda\to0$,
the virial coefficient, computed in Eq.~(\ref{v2}), tends to $\pi/2$
(for positive mass) as expected.  When $\lambda\to1$, on the other
hands, with the asymptotics of $v_2$ found in Eq.~(\ref{v2}),
Eq.~(\ref{idealGas}) can be rewritten as
\begin{equation}
  \frac p{nT} = n - \frac\pi 2 \frac n{N_{\rm boson}m_0 T}\,,
\end{equation}
in complete agreement with the interpretation of the system as an
almost ideal $N_{\rm boson}$-component Bose gas.

We now take the bosons to temperatures below the boson degeneracy
temperature $T_{\rm bos.deg}$.  Recall that in two spatial dimensions,
noninteracting bosons do not form Bose-Einstein condensate: as one
lowers the temperature the chemical potential approaches 0 from below,
but never reaches 0.  For $T\ll T_{\rm bos.deg.}$, when $\mu$ is close
to 0, the energy density of a free Bose gas is
\begin{equation}
  \epsilon = N_{\rm boson} \int\! \frac{d^2{\bf k}}{(2\pi)^2}\,
  \frac{\epsilon_{\bf k}}{e^{\epsilon_{\bf k}/T}-1}
  = N_{\rm boson} \frac{\pi}{12}mT^2 
\end{equation}
(here $\epsilon_{\bf k}=\frac{k^2}{2m}$),
and hence the specific heat in the boson degeneracy regime is
\begin{equation}
  c_v = N_{\rm boson}\frac{\pi}6 mT = (1-\lambda)N \frac{\pi}6 mT\,,
\end{equation}
which matches exactly the linear slope found numerically in the
previous section.

When $T\lesssim T_q = (1-\lambda)T_{\rm bos.deg.}$ our calculations
indicate that the system is no longer an ideal Bose gas.  Presumably,
at these low temperatures the interactions between the bosons can no
longer be ignored.  That the bosonic system turns into a Fermi liquid
at very low temperature is a miracle of Bose-Fermi duality.

\section{The Landau Parameters}\label{sec:LandauParameters}

In this section we argue that a large-$N$ Chern-Simons system with finite density fermions
(\ref{FullLagrangian}) behaves as a Landau Fermi liquid at low temperatures.
To support this statement we calculate the Landau parameters of the
theory (\ref{FullLagrangian}) by two different methods, thermodynamic and microscopic, and demonstrate
that in both cases we obtain the same result, describing a stable, interacting
Fermi liquid.
The thermodynamic method relies on the equation of state (\ref{FreeEnergy}) derived in
section \ref{sec:Thermo}. We use the free energy (\ref{FreeEnergy}) to work out quasiparticle effective mass and compressibility of the system. Matching these to the predictions of LFL theory,
we can derive the values of the Landau parameters.

Then we work out the Landau parameters directly using
the microscopic formula (\ref{InteractionFunction}). This is made possible by the large $N$ limit and our gauge choice, which restricts the type of diagrams that contribute at leading order in $N$ and allow one to write down an exact Schwinger-Dyson equation for scattering amplitudes. The integral equations we shall need for the vertex function were first given in section 5.2 of \cite{Giombi:2011kc} and latter solved in \cite{Jain:2014nza,Inbasekar:2015tsa} at zero chemical potential. Prior to this, similar calculations of two point correlators were performed in \cite{Aharony:2012nh,GurAri:2012is} and our analysis in sections \ref{sec:Conductivity} and \ref{sec:Viscosity} will closely follow these references. To calculate the Landau parameters we shall need the solution for nonzero $\mu$, evaluated at the Fermi surface. The calculation proceeds essentially as those found in these references and is not particularly instructive. The interested reader may find the details in appendix \ref{app:4ptVertexCalc}.

\subsection{Thermodynamic Calculation of the Landau Parameters}
\label{sec:LandauParametersfromThermodynamics}

In section \ref{sec:Thermo} we found the low temperature equation of state (\ref{FreeEnergy}),
and used it to derive the entropy density and heat capacity (\ref{LowTempcV}).
We also worked out the expression for entropy density by a direct statistical calculation in (\ref{CvLowTemp})
for a system of quasiparticles with the effective mass $m^\star$. Comparing these
two results for the heat capacity we see that the effective mass is simply given by
\begin{align}
	m^\star = \mu \,.
\end{align}
Also note that this formula may also be extracted from the low temperature limit of the Green's function (\ref{FermionPropagator})
expanded near the Fermi surface and matched against the LFL
quasiparticle propagator (\ref{LFLFermionicPropagator}). From the
expression (\ref{effective mass}) we see that the exchange channel makes a subleading contribution in the large $N$ limit, while the Landau parameter
in the direct channel vanishes
\begin{align}
	F^{(d)}_1 = 0 \,.
\end{align}

Now let us turn to the zeroth Landau parameters, related to the compressibility by (\ref{compressibility}).
The isothermal inverse compressibility is simply calculated from the low temperature equation of state (\ref{FreeEnergy})
\begin{align}
	\kappa^{-1} = n^2 \left( \frac{\p \mu}{\p n} \right)_T = \frac{2 \pi n^2 }{N ( \mu - \lambda c_0)} .
\end{align}
Again the exchange channel does not contribute in the large $N$ limit and comparing to (\ref{compressibility}), we have
\begin{equation}\label{F0}
	F^{(d)}_0 =\frac{\lambda\,c_0}{\mu-\lambda\,c_0}\,.
\end{equation}
This result can also be obtained in the massless case $m=0$ by demanding the speed
of zero sound take the conformal value $s=1/\sqrt{2}$.

\subsection{The Four-point Vertex Function}\label{sec:4ptVertex}
We now turn to a direct calculation of the Landau parameters from the definition (\ref{InteractionFunction}). We begin by evaluating the 1PI four-point function.
In the large $N$ limit, the four-point vertex function (\ref{V def}) is a sum of ladder diagrams and so obeys a Schwinger-Dyson integral equation. In the direct channel this is\footnote{The factors of $1/2$ in front of both terms in the r.h.s.
originate from the normalization convention of the gauge propagator (\ref{GaugePropagator}),
chosen to match \cite{Giombi:2011kc}.}
\begin{align}
\label{RecursionRelation}	
V^{(d)} (p , k , q )^\alpha{}_\delta,{}^\gamma{}_\beta 
&=-  \frac{1}{2} G_{\mu \nu} (p-k) (\gamma^\mu)^\alpha{}_\beta (\gamma^\nu)^\gamma{}_\delta \\
	&- \frac{1}{2} \int \frac{d^3 r}{(2 \pi)^3} G_{\mu \nu} (p - r)  \left( \gamma^\mu G (r+q) V^{(d)} ( r , k , q ),{}^\gamma{}_\beta G ( r ) \gamma^\nu \right)^\alpha{}_\delta \nonumber
\end{align}
or, diagrammatically (in the second diagram in the r.h.s. the internal fermionic lines with momenta
$r$ and $r+q$ are the full fermionic propagators, which we draw as simple lines not to clutter the picture)
\begin{center}
\fcolorbox{white}{white}{
   \scalebox{.8}{
  \begin{picture}(529,108) (16,-73)
    \SetWidth{1.3}
    \SetColor{Black}
    \Line[arrow,arrowpos=0.5,arrowlength=8.8,arrowwidth=3.3,arrowinset=0.2](118,8)(160,8)
    \Line[arrow,arrowpos=0.5,arrowlength=8.8,arrowwidth=3.3,arrowinset=0.2](161,-52)(121,-52)
    \SetWidth{1.0}
    \GOval(98,-22)(50,24)(0){0.882}
    \SetWidth{1.3}
    \Text(195,-27)[lb]{{\Black{$=$}}}
    \Text(325,-27)[lb]{{\Black{$+$}}}
    \Line[arrow,arrowpos=0.5,arrowlength=8.8,arrowwidth=3.3,arrowinset=0.2](42,8)(78,8)
    \Line[arrow,arrowpos=0.5,arrowlength=8.8,arrowwidth=3.3,arrowinset=0.2](78,-52)(42,-52)
    \Text(13,4)[lb]{{\Black{$\alpha\;\; i$}}}
    \Text(47,15)[lb]{{\Black{$p+q$}}}
    \Text(123,15)[lb]{{\Black{$k+q$}}}
    \Text(163,4)[lb]{{\Black{$\beta\;\; j$}}}
    \Text(13,-60)[lb]{{\Black{$\delta \;\; l$}}}
    \Text(56,-70)[lb]{{\Black{$p$}}}
    \Text(137,-70)[lb]{{\Black{$k$}}}
    \Text(163,-60)[lb]{{\Black{$\gamma\;\; k$}}}
    \Line[arrow,arrowpos=0.5,arrowlength=8.8,arrowwidth=3.3,arrowinset=0.2](477,8)(519,8)
    \Line[arrow,arrowpos=0.5,arrowlength=8.8,arrowwidth=3.3,arrowinset=0.2](518,-52)(478,-52)
    \SetWidth{1.0}
    \GOval(456,-22)(50,24)(0){0.882}
    \SetWidth{1.3}
    \Line[arrow,arrowpos=0.5,arrowlength=8.8,arrowwidth=3.3,arrowinset=0.2](399,8)(435,8)
    \Line[arrow,arrowpos=0.5,arrowlength=8.8,arrowwidth=3.3,arrowinset=0.2](436,-52)(400,-52)
    \Line[arrow,arrowpos=0.5,arrowlength=8.8,arrowwidth=3.3,arrowinset=0.2](399,-52)(399,7)
    \Line[arrow,arrowpos=0.5,arrowlength=8,arrowwidth=3,arrowinset=0.2](391,8)(391,-52)
    \Line[arrow,arrowpos=0.5,arrowlength=8,arrowwidth=3,arrowinset=0.2](391,-52)(361,-52)
    \Line[arrow,arrowpos=0.5,arrowlength=8,arrowwidth=3,arrowinset=0.2](361,8)(393,8)
    \Text(204,4)[lb]{{\Black{$\alpha\;\; i$}}}
    \Text(230,15)[lb]{{\Black{$p+q$}}}
    \Text(275,15)[lb]{{\Black{$k+q$}}}
    \Text(308,4)[lb]{{\Black{$\beta\;\; j$}}}
    \Text(204,-60)[lb]{{\Black{$\delta \;\; l$}}}
    \Text(275,-70)[lb]{{\Black{$k$}}}
    \Text(239,-70)[lb]{{\Black{$p$}}}
    \Text(308,-60)[lb]{{\Black{$\gamma\;\; k$}}}
    \Text(220,-27)[lb]{{\Black{$p-k$}}}
    \Line[arrow,arrowpos=0.5,arrowlength=8.8,arrowwidth=3.3,arrowinset=0.2](267,8)(303,8)
    \Line[arrow,arrowpos=0.5,arrowlength=8.8,arrowwidth=3.3,arrowinset=0.2](302,-52)(266,-52)
    \Line[arrow,arrowpos=0.5,arrowlength=8.8,arrowwidth=3.3,arrowinset=0.2](267,-52)(267,7)
    \Line[arrow,arrowpos=0.5,arrowlength=8,arrowwidth=3,arrowinset=0.2](261,8)(261,-52)
    \Line[arrow,arrowpos=0.5,arrowlength=8,arrowwidth=3,arrowinset=0.2](261,-52)(231,-52)
    \Line[arrow,arrowpos=0.5,arrowlength=8,arrowwidth=3,arrowinset=0.2](230,8)(262,8)
    \Text(338,4)[lb]{{\Black{$\alpha\;\; i$}}}
    \Text(362,15)[lb]{{\Black{$p+q$}}}
    \Text(402,15)[lb]{{\Black{$r+q$}}}
    \Text(338,-60)[lb]{{\Black{$\delta \;\; l$}}}
    \Text(371,-70)[lb]{{\Black{$p$}}}
    \Text(411,-70)[lb]{{\Black{$r$}}}
    \Text(350,-27)[lb]{{\Black{$p-r$}}}
     \Text(483,15)[lb]{{\Black{$k+q$}}}
    \Text(523,4)[lb]{{\Black{$\beta\;\; j$}}}
     \Text(483,-70)[lb]{{\Black{$k$}}}
    \Text(523,-60)[lb]{{\Black{$\gamma\;\; k$}}}
  \end{picture}
  }
}
\end{center}

In equation (\ref{RecursionRelation}) we have suppressed color indices as their placement is encoded in 't Hooft's double line notation. 
The notation in the integrand indicates that spinor indices are contracted as if matrices are being multiplied in the indicated order.
Note that in (\ref{RecursionRelation}) we have organized the spin indices on the vertex function into two factors, separated by a comma. This is convenient since only the first factor participates in the matrix multiplication of the final term. The second factor is then simply along for the ride.

Setting $q_s = 0$, this integral equation can be solved for any value of the incoming momenta $p$ and $k$. For our purposes however, we only require the solution on the Fermi surface, where $p_s = k_s = p_F$. The particles with momenta $p$ and $k$ are incident with angles $\theta_p$ and $\theta_k$ respectively (by rotational invariance our answer can depend only on $\theta = \theta_p - \theta_k$). We have also checked up to one-loop order that the perturbative calculation agrees with an
expansion of the full
answer around $\lambda=0$.
Here we
simply state the result
\begin{align}\label{4ptVertex}
	V^{(d)}(\theta_p, \theta_k)
	=\frac{2\pi\lambda^2}{\mu - \lambda c_0} I \otimes I
	+\frac{2\sqrt{2}\pi i\lambda}{p_F(e^{i\theta_k}-e^{i\theta_p})}\left( I \otimes \gamma^+ - \gamma^+ \otimes I
	\right)\,,
\end{align}
where we have taken the exchange momentum $q \rightarrow 0$ in the rapid limit:
first $q_s\rightarrow 0$, second $q_3\rightarrow 0$. The product decomposition of matrices corresponds to our organization of spin indices in $V^{(d)\alpha}{}_\delta{},^\gamma{}_\beta$. That is, the first matrix corresponds to indices to the left of the comma and the second matrix to those on the right.

\subsection{Microscopic Calculation of the Landau Parameters}\label{sec:LandauCalc}
To complete our calculation of the Landau parameters we need to work out the quasiparticle spinor $u(p)$ and wave function renormalization $Z$.
As reviewed in section \ref{sec:CSintro}, the large $N$ fermionic propagator is known and is given by equations (\ref{FermionPropagator}), (\ref{SelfEnergyMatrix}) and (\ref{SelfEnergy}).
The full dressed quasi-particle spinor then satisfies
\begin{equation}
	\label{DiracEquation}
	\left( i\gamma^\mu \tilde p_\mu + \Sigma(p) \right)_\alpha{}^\beta u_\beta (p)=0 .
\end{equation}

We solve this on the Fermi surface. Here the self-energy (\ref{SelfEnergy}) is simply $\Sigma=c_0 I$,
and the momentum components on Fermi surface are
$p_\pm =\frac{1}{\sqrt{2}}p_Fe^{\mp i\theta}$ and $\tilde p_3=i\mu$. The Dirac equation
then assumes the form
\begin{align}
	\label{DiracEquationFermiSurface}
	\begin{pmatrix}
		c_0 - \mu & i p_F e^{- i \theta} \\
		i p_F e^{i \theta} & c_0 + \mu
	\end{pmatrix}
	\begin{pmatrix}
		u_1 \\
		u_2
	\end{pmatrix}
	= 0 .
\end{align}
The solution describing quasiparticles is
\begin{equation}
	\label{UonFermiSurface}
	u_\alpha =
	\left({\sqrt{\mu+c_0}\,e^{-i\theta/2}\atop \,-i \sqrt{\mu-c_0}\,e^{i\theta/2}}\right)\,,
\end{equation}
where we have imposed the normalization $u^\dagger u=2\mu$.
Expanding the Green's function (\ref{FermionPropagator}) about the Fermi-surface
and matching with the Landau Fermi liquid expression (\ref{LFLFermionicPropagator})
we find that $Z=1$.
 Assembling this all together we may finally compute the Landau parameters using (\ref{V def}) and (\ref{InteractionFunction})\footnote{This formula was written in Lorentzian signature, while this section has been in Euclidean signature. To use (\ref{UonFermiSurface}) in (\ref{InteractionFunction}) one must recall that $\bar u = \sigma^3 u^*$ in Lorentzian space.}. Remarkably, all angular dependence drops out and the interaction strength is constant along the Fermi surface. There is then only the single non-zero parameter in the direct channel
\begin{equation}
	F^{(d)}_0 =\frac{\lambda\,c_0}{\mu-\lambda\,c_0}\,,
\end{equation}
which in agreement with (\ref{F0}). 

We do not calculate the Landau parameters in the exchange channel since this requires solving the recursion relation at all values of $q_\pm$. This can be easily seen from the recursion relation in the exchange channel, in which $q$ does not simply appear as a parameter
\begin{align}
	V^{(e)} ( p ; k ; q ) &= -\frac 1 2 G_{+3} (q) \left( \gamma^+ \otimes I - I \otimes \gamma^+ \right) \nonumber \\
		&- \frac 1 2 \int \frac{d^3r}{(2 \pi)^3} G_{+3}(r-q) H_+ \left( G (p+q ) V^{(e)} ( p , k ; r ) G (k + r) \right) .
\end{align}
A similar difficulty prevents a direct evaluation of the $S$ matrix in the $S$-channel in \cite{Jain:2014nza}. One can obtain the vertex function in the exchange channel from the direct channel by use of Fermi statistics, but this requires knowledge of $V^{(d)} (p,k;q)$ at finite $q_\pm$. Since we lack Lorentz invariance in the presence of a Fermi surface we cannot infer this from our results above. Although we cannot compute the exchange Landau parameters directly, we can still perform the comparison above since the contribution of the exchange channel is subleading in $N$ for the observables (\ref{effective mass}) and (\ref{compressibility}) we have considered.

\section{The Conductivity Tensor}\label{sec:Conductivity}
In this section we evaluate the conductivity tensor at large $N$ for any frequency $\omega$; to our knowledge, the first exact evaluation within an interacting field theory for all values of the coupling constant. We draw particular attention to the zero frequency Hall conductivity, which is not simply the proportional to the total Berry flux (\ref{HallConductivity}). In future work we will pursue what is needed to completely capture the Hall conductivity within Fermi liquid theory \cite{Jingyuan2015}.

We calculate the conductivity tensor by its Kubo formula
\begin{equation}
\label{Kubo}
\sigma^{ij}(\omega)=\frac{1}{i \omega_+} G^{i,j}_R (\omega),
\end{equation}
where $G^{i,j}_R (\omega)$ is the Fourier transformed retarded Green's function\footnote{We have set the contact term to zero since (\ref{chargecurrent}) has no $A_\mu$ dependence.}
\begin{align}
	G^{i,j}_R ( \omega ) &=  i \int_{- \infty}^\infty d^3 x e^{i \omega_+ x^0} \Theta ( x^0 )\left \langle [ j^{i} (x) , j^{j} (0) ] \right \rangle  .
\end{align}
and
\begin{equation}
j^\mu =i\,\bar\psi \gamma^\mu \psi\,
\label{chargecurrent}
\end{equation}
is the $U(1)$ current. The frequency is always evaluated with a small, positive imaginary part $\omega_+ = \omega + i 0^+$ which we will simply denote as $\omega$ from this point forward. This calculation was performed at zero density and zero mass in \cite{GurAri:2012is}. We will first calculate the Euclidean time ordered correlators and obtain the retarted Green's functions by Wick rotating back to Minkowski space $q_3 \rightarrow i \omega$.

Let's start by evaluating
the three-point vertex function
involving the current and two fermions
\begin{center}
\fcolorbox{white}{white}{
  \scalebox{.5}{
  \begin{picture}(248,248) (115,-27)
    \SetWidth{1.0}
    \SetColor{Black}
    \GOval(232,80)(40,40)(0){0.882}
    \Text(220,75)[lb]{\Large{\Black{${\bf 1\;  PI}$}}}
    \Line[arrow,arrowpos=0.5,arrowlength=8,arrowwidth=3,arrowinset=0.2](144,192)(208,112)
    \Line[arrow,arrowpos=0.5,arrowlength=8,arrowwidth=3,arrowinset=0.2](208,48)(144,-24)
    \Photon(272,80)(360,80){7.5}{4}
    \scalebox{1.3}{
    \Text(-30,60)[lb]{\Large{\Black{$V^{\mu,\alpha}{}_\beta (p ; q )\quad =$}}}
    }
    \Text(184,160)[lb]{\Large{\Black{$p+q$}}}
    \Text(192,8)[lb]{\Large{\Black{$p$}}}
    \Text(125,-35)[lb]{\Large{\Black{$\beta$}}}
    \Text(125,190)[lb]{\Large{\Black{$\alpha$}}}
    \Text(310,96)[lb]{\Large{\Black{$q$}}}
    \Text(370,75)[lb]{\Large{\Black{$\mu$}}}
  \end{picture}
  }
}
\end{center}
from which we then find the current-current correlator
\begin{center}
\fcolorbox{white}{white}{
  \scalebox{.5}{
    \begin{picture}(200,208) (83,-27)
    \SetWidth{1.0}
    \SetColor{Black}
       \scalebox{1.3}{
    \Text(80,40)[lb]{\Large{\Black{$ \left \langle j^i (q) j^j (-q) \right \rangle 
	 =$}}}
    }
      \scalebox{1.3}{
    \Text(400,40)[lb]{\Large{\Black{$ =iN \int \frac{d^3 p}{(2 \pi )^3} \text{Tr} ( V^j (p;q) G(p+q) \gamma^i   G(p) ).$}}}
    }
     \end{picture}
  \begin{picture}(976,208) (83,-27)
    \SetWidth{1.0}
    \SetColor{Black}
    \GOval(304,64)(19,19)(0){0.882}
    \GOval(192,128)(16,16)(0){0.882}
    \GOval(192,0)(16,16)(0){0.882}
    \Vertex(96,64){7.071}
    \Arc[arrow,arrowpos=0.5,arrowlength=5,arrowwidth=2,arrowinset=0.2,clock](178.667,42.667)(85.375,165.53,91.79)
    \Arc[arrow,arrowpos=0.5,arrowlength=5,arrowwidth=2,arrowinset=0.2,clock](178.667,85.333)(85.375,-91.79,-165.53)
    \Arc[arrow,arrowpos=0.5,arrowlength=5,arrowwidth=2,arrowinset=0.2,clock](216,24)(104.307,94.399,32.471)
    \Arc[arrow,arrowpos=0.5,arrowlength=5,arrowwidth=2,arrowinset=0.2,clock](216,104)(104.307,-32.471,-94.399)
    \Text(70,80)[lb]{\Large{\Black{$j$}}}
    \Text(176,160)[lb]{\Large{\Black{$p+q$}}}
    \Text(192,-32)[lb]{\Large{\Black{$p$}}}
    \Text(330,80)[lb]{\Large{\Black{$i$}}}
 
  \end{picture}
  }
}
\end{center}

The vertex $V^{\mu,\alpha}{}_\beta$ in the large $N$ limit obeys the recursion relation
\begin{align}
\label{vertexrecursionrelation}
	V^\mu (p;q) = i  \gamma^\mu - \frac 1 2
	\int \frac{ d^3 r}{(2 \pi)^3} G_{\rho \sigma} ( p - r ) \gamma^\rho G(r+q) V^\mu(r;q) G(r) \gamma^\sigma ,
\end{align}
which diagramatically may be expressed as
\begin{center}
\fcolorbox{white}{white}{
  \scalebox{.8}{
  \begin{picture}(534,149) (67,-53)
    \SetWidth{0.5}
    \SetColor{Black}
    \Text(304,57)[lb]{{\Black{$p+q$}}}
    \SetWidth{1.0}
    \Line[arrow,arrowpos=0.5,arrowlength=8,arrowwidth=3,arrowinset=0.2](82,70)(129,29)
    \GOval(135,11)(18,18)(0){0.882}
    \Line[arrow,arrowpos=0.5,arrowlength=8,arrowwidth=3,arrowinset=0.2](128,-7)(81,-48)
    \Photon(152,11)(205,11){6}{3}
    \Text(233,6)[lb]{{\Black{$=$}}}
    \Line[arrow,arrowpos=0.5,arrowlength=8,arrowwidth=3,arrowinset=0.2](274,70)(333,11)
    \Line[arrow,arrowpos=0.5,arrowlength=8,arrowwidth=3,arrowinset=0.2](332,11)(274,-48)
    \Vertex(332,11){5}
    \Text(368,6)[lb]{{\Black{$+$}}}
    \Line[arrow,arrowpos=0.5,arrowlength=8,arrowwidth=3,arrowinset=0.2](391,70)(432,70)
    \Line[arrow,arrowpos=0.5,arrowlength=8,arrowwidth=3,arrowinset=0.2](432,70)(432,-47)
    \Line[arrow,arrowpos=0.5,arrowlength=8,arrowwidth=3,arrowinset=0.2](432,-47)(392,-47)
    \Line[arrow,arrowpos=0.5,arrowlength=8,arrowwidth=3,arrowinset=0.2](438,-47)(438,70)
    \Line[arrow,arrowpos=0.5,arrowlength=8,arrowwidth=3,arrowinset=0.2](438,70)(502,23)
    \Line[arrow,arrowpos=0.5,arrowlength=8,arrowwidth=3,arrowinset=0.2](502,0)(438,-47)
    \GOval(508,11)(18,18)(0){0.882}
    \Photon(525,11)(578,11){6}{3}
    \Text(67,70)[lb]{{\Black{$\alpha$}}}
    \Text(67,-47)[lb]{{\Black{$\beta$}}}
    \Text(110,57)[lb]{{\Black{$p+q$}}}
    \Text(120,-34)[lb]{{\Black{$p$}}}
    \Text(176,23)[lb]{{\Black{$q$}}}
    \Text(215,6)[lb]{{\Black{$\mu$}}}
    \Text(260,70)[lb]{{\Black{$\alpha$}}}
    \Text(260,-47)[lb]{{\Black{$\beta$}}}
    \Text(310,-34)[lb]{{\Black{$p$}}}
    \Text(380,70)[lb]{{\Black{$\alpha$}}}
    \Text(380,-47)[lb]{{\Black{$\beta$}}}
    \Text(400,80)[lb]{{\Black{$p+q$}}}
    \Text(410,-68)[lb]{{\Black{$p$}}}
    \Text(390,6)[lb]{{\Black{$p-r$}}}
    \Text(473,52)[lb]{{\Black{$r+q$}}}
    \Text(473,-39)[lb]{{\Black{$r$}}}
    \Text(548,23)[lb]{{\Black{$q$}}}
    \Text(566,-7)[lb]{{\Black{$\mu$}}}
  \end{picture}
  }
}
\end{center}

As before, solving (\ref{vertexrecursionrelation}) is rather cumbersome. The details are collected in appendix \ref{app:currentvertex}. In the end, the only independent nonzero correlator is
\begin{align}
\label{jpjm}
\left \langle j^+ (\omega ) j^- ( - \omega ) \right\rangle=  - \frac{N}{16 \pi \lambda \omega}\left( (\omega - 2 c_0)^2 \left( 1 - e^{2 \lambda {\rm arctanh} \frac{\omega}{2 \mu}} \right)+ 4 \lambda \mu \omega \right) ,
\end{align}
where we have introduced counterterms to subtract out a linear divergence. The 
correlation function $\left \langle j^- (\omega ) j^+ ( - \omega )\right\rangle$ is obtained from (\ref{jpjm}) by the replacing $\omega\rightarrow -\omega$.

The longitudinal and Hall conductivities are then simply
\begin{align}\label{condDecomp}
	&\sigma  (\omega) = \frac{1}{2} \delta_{ij} \sigma^{ij} (\omega) = \frac{1}{2} ( \sigma^{+-}  (\omega) + \sigma^{+ -}  (- \omega) ) ,\nonumber \\
	&\sigma_H  (\omega) = \frac{1}{2} \epsilon_{ij} \sigma^{ij} = \frac{i}{2 } ( \sigma^{+-}  (\omega) - \sigma^{+ -}  (-\omega) ) .
\end{align}
Altogether then, we have
\begin{align}
	\sigma (\omega ) &= - \frac{N i}{32 \pi \lambda \omega^2} \left( 8 \omega ( c_0 - \lambda \mu ) + ( \omega - 2 c_0 )^2 e^{2 \lambda \arctanh \frac {\omega}{2 \mu}}- ( \omega + 2 c_0 )^2 e^{-2 \lambda \arctanh \frac {\omega}{2 \mu}} \right) ,\nonumber \\
	\sigma_H (\omega) &= - \frac{N}{32 \pi \lambda \omega^2} \left( -2 ( \omega^2 + 4 c_0^2 ) + ( \omega - 2 c_0 )^2 e^{2 \lambda \arctanh \frac {\omega}{2 \mu}}+ ( \omega + 2 c_0 )^2 e^{-2 \lambda \arctanh \frac {\omega}{2 \mu}} \right) .
\end{align}

\begin{figure}
\begin{center}
\includegraphics[width=.45\textwidth]{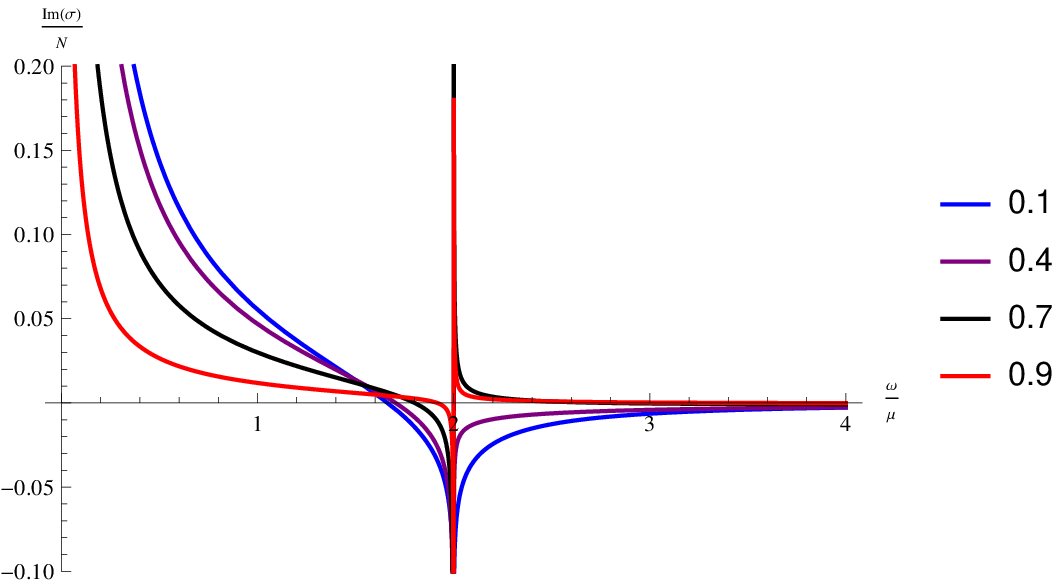}
\includegraphics[width=.45\textwidth]{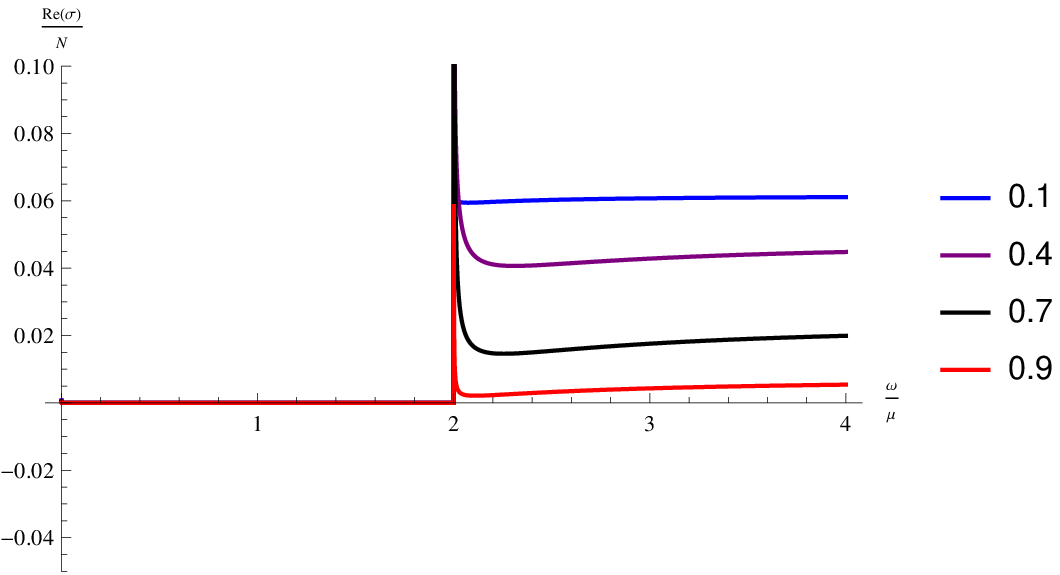}
\end{center}
\caption{Longitudinal conductivity $\sigma$ at $m_0 / \mu = .1$ and several values of $\lambda$.}
\label{fig:LongitudinalConductivity}
\end{figure}

\begin{figure}
\begin{center}
\includegraphics[width=.45\textwidth]{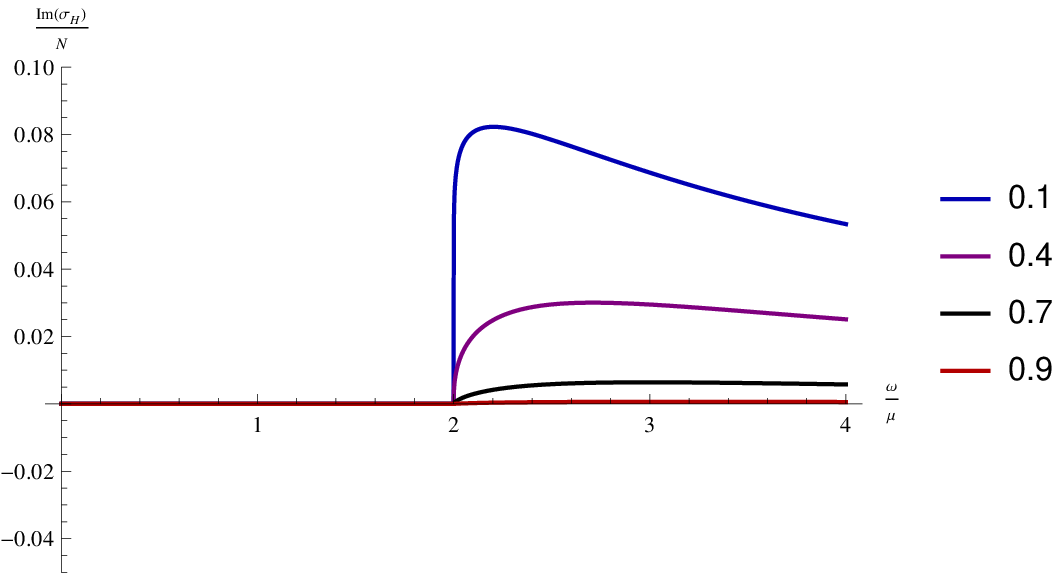}
\includegraphics[width=.45\textwidth]{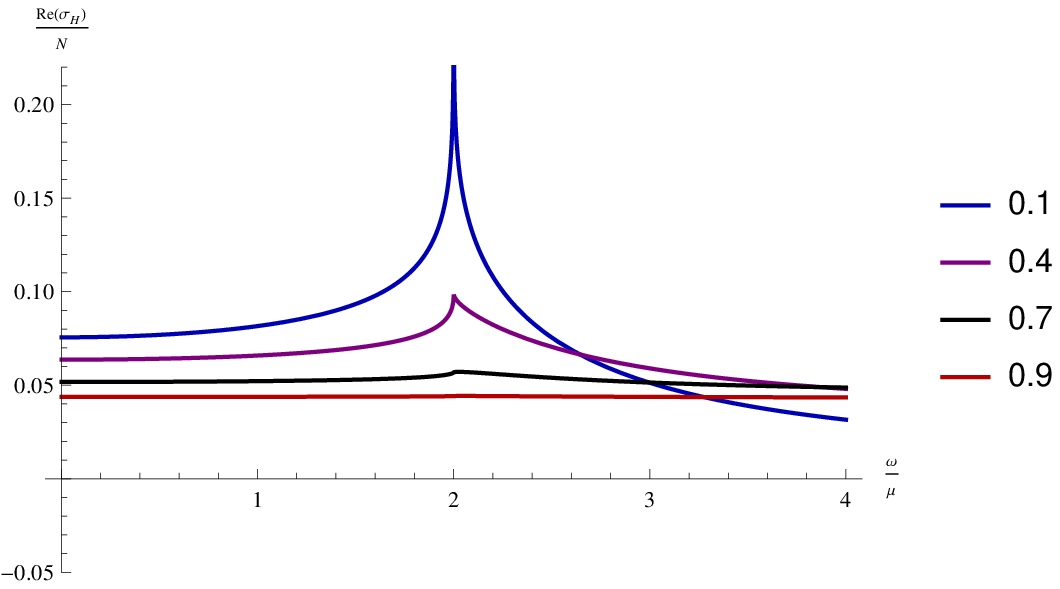}
\end{center}
\caption{Hall conductivity $\sigma_H$ at $m_0 / \mu = 1$ and several values of $\lambda$.}
\label{fig:HallConductivity}
\end{figure}

These may be found plotted in figures \ref{fig:LongitudinalConductivity},
\ref{fig:HallConductivity}. The discontinuous feature at $\omega = 2 \mu$ arises from a branch cut in the Green's function signifying a continuum of multi-particle states beginning at the pair-production threshold. The real part of the dissipative conductivity is zero below this threshold, while its imaginary part diverges as $1/\omega$ in accordance with the Drude formula (\ref{drude}).

It is instructive to consider conductivity in several different regimes. At frequencies much larger than any other scales, we retrieve the results of \cite{GurAri:2012is}
\begin{align}
	\sigma ( \omega ) =  \frac{N \sin \pi \lambda}{16 \pi \lambda} + \mathcal O ( \omega^{-1}),
	&&
	\sigma_H ( \omega ) =-  \frac{N \sin^2 \frac{\pi \lambda}{2}}{8 \pi \lambda} + \mathcal O ( \omega^{-1}).
\end{align}
The Fermi liquid description on the other hand is valid at low frequencies $\omega \rightarrow 0$
\begin{align}\label{drude}
	\sigma =  \frac{N p_F^2}{4 \pi \mu} \frac{1}{- i \omega} + \mathcal O ( \omega  ),
	&&\sigma_H =- \frac{N c_0 }{4 \pi \mu^2}  \left( \mu - \frac 1 2 \lambda c_0  \right) + \mathcal O ( \omega^2 ).
\end{align}
The longitudinal conductivity simply reduces to the Drude formula. Recall that since $\omega$ has a small, positive imaginary part, the real part of the dissipative conductivity also includes a delta function at zero frequency $\sigma = \frac{\pi n}{\mu} \delta (\omega ) + \cdots$ that is implicit in our formulas above. We expect that this would be broadened at finite $N$ by quasi-particle decay (see comments below (\ref{gapEquations})).

It is easy to see that the Hall conductivity does not match our expectations from (\ref{HallConductivity}). Using the wavefunctions at the Fermi surface (\ref{UonFermiSurface}), the enclosed Berry flux is simply
\begin{align}
	\frac{1}{4\pi^2} \oint_{p_F} \Tr ~ \mathcal A = - \frac{N c_0 }{4 \pi \mu}  
\end{align}
and so does not entirely account for the Hall conductivity of the Fermi liquid state when interactions are present.

\section{The Viscosity Tensor}\label{sec:Viscosity}
We now undertake a similar analysis of the viscosity tensor. The viscosity encodes the stress induced by shearing within linear response theory
\begin{align}\label{linearResponseStress}
	\left \langle T^{ij} \right \rangle = - p \delta^{ij} + \eta^{ijkl} \tau_{ij} + \cdots ,
	&&\text{where}
	&&\tau_{ij} = \partial_i u_j + \partial_j u_i 
\end{align}
and $u_i$ is the local fluid velocity.
After a time-dependent diffeomorphism to the fluid Lagrangian coordinates, this appears as the response of the stress tensor to deformations of the fluid internal metric $\tau_{ij} = i \omega_+ g_{ij}$, and so may be captured by a stress-tensor Kubo formula. For a comprehensive treatment of viscosity within linear response theory we refer the reader to \cite{Bradlyn2012}. In 2+1 dimensions, $\eta^{ijkl}$ has three independent components, the bulk, shear and Hall viscosities respectively \cite{AvronPRL}
\begin{align}
	\eta^{ijkl} = \zeta \delta^{ij} \delta^{kl} + \eta \Pi^{ijkl} + \tilde \eta \tilde \Pi^{ijkl} .
\end{align}
where we have introduced the even and odd projectors
\begin{align}
	&\Pi^{ijkl} = \delta^{i(k} \delta^{l)j} - \frac 1 2 \delta^{ij} \delta^{kl},
	&&\tilde \Pi^{ijkl} = \frac 1 2 \left( \delta^{i(k} \epsilon^{l)j} + \delta^{j (k} \epsilon^{l)i} \right) .
\end{align} 
Their Kubo formulas are then
\begin{align} 
	\zeta ( \omega ) &= - \frac{1}{4 i \omega_+} \delta_{ij} \delta_{kl} G^{ij,kl}_R ( \omega ) 
	+ \frac{ p + \kappa^{-1}}{2 i \omega_+} , \nonumber \\
	\eta ( \omega ) &= - \frac{1}{ 2 i \omega_+} \Pi_{ijkl} G^{ij,kl}_R ( \omega ) 
	+ \frac{p}{ i \omega_+} ,\label{viscousKubo} \\
	\eta_H ( \omega ) &= - \frac{1}{ 2 i \omega_+} \tilde \Pi_{ijkl} G^{ij,kl}_R ( \omega ) ,\nonumber
\end{align}
where we have denoted the Fourier transformed retarded Green's function, including contact terms, as
\begin{align}\label{GijklFormula}
	G^{ij,kl}_R ( \omega ) &=  \int_{- \infty}^\infty d^3 x e^{i \omega_+ x^0} \left( \left \langle \frac{\delta T^{ij} (x) }{\delta g_{kl} (0)}\right \rangle + \frac{i}{2} \Theta ( x^0 )\left \langle [ T^{ij} (x) , T^{kl} (0) ] \right \rangle \right) .
\end{align}
As pointed out in \cite{Bradlyn2012}, the thermodynamic terms appearing in (\ref{viscousKubo}) are necessary to subtract off contributions from the first term of (\ref{linearResponseStress}) under metric perturbations.

The stress tensor of our theory is given by
\begin{align}
	T^{\mu \nu} = - \frac 1 2 \bar \psi \gamma^{(\mu} \overset{\leftrightarrow} D{}^{\nu)} \psi+ \left( \frac{1}{2} \bar \psi \gamma^\lambda \overset{\leftrightarrow} D_\lambda \psi + m \bar \psi \psi \right) \eta^{\mu \nu} .
\end{align}
The cosmological constant introduced to cancel the vacuum energy density in (\ref{EOS}) should also appear here and is necessary to get the correct pressure in (\ref{linearResponseStress}). However, we can safely ignore this in a viscosity computation as the pressure and compressibility terms in the Kubo formulas (\ref{viscousKubo}) are introduced so as to make the viscosities independent of the equation of state, and one can easily verify that a cosmological constant in particular does not affect them.

As before, the first step is to calculate the vertex function with a single stress insertion
\begin{center}
\fcolorbox{white}{white}{
  \scalebox{.5}{
  \begin{picture}(248,248) (115,-27)
    \SetWidth{1.0}
    \SetColor{Black}
    \GOval(232,80)(40,40)(0){0.882}
    \Text(220,75)[lb]{\Large{\Black{${\bf 1\;  PI}$}}}
    \Line[arrow,arrowpos=0.5,arrowlength=8,arrowwidth=3,arrowinset=0.2](144,192)(208,112)
    \Line[arrow,arrowpos=0.5,arrowlength=8,arrowwidth=3,arrowinset=0.2](208,48)(144,-24)
    \Photon(272,80)(360,80){7.5}{4}
    \scalebox{1.3}{
    \Text(-30,60)[lb]{\Large{\Black{$N U^{\mu\nu,\alpha}{}_\beta (p;q)\quad =$}}}
    }
    \Text(184,160)[lb]{\Large{\Black{$p+q$}}}
    \Text(192,8)[lb]{\Large{\Black{$p$}}}
    \Text(125,-35)[lb]{\Large{\Black{$\beta$}}}
    \Text(125,190)[lb]{\Large{\Black{$\alpha$}}}
    \Text(310,96)[lb]{\Large{\Black{$q$}}}
    \Text(370,75)[lb]{\Large{\Black{$\mu\nu$}}}
  \end{picture}
  }
}
\end{center}
which is a three-point function
\begin{equation}
\left( U^{\mu\nu}\right)^\alpha{}_\beta (p,q)=
\left\langle T^{\mu\nu}(q)\psi _\beta (p)
\bar\psi ^\alpha (- p - q)\right\rangle_\text{1PI}\,.
\end{equation}
In the large $N$ limit it obeys the recursion relation
\begin{align}
\label{stressvertexrecursionrelation}
	U^{\mu\nu} (p,q) = 
	U_0^{\mu\nu} - \frac 1 2
	\int \frac{ d^3 r}{(2 \pi)^3} G_{\rho \sigma} ( p - r) \gamma^\rho G(r+q)
	 U^{\mu\nu} (r;q) G(r) \gamma^\sigma ,
\end{align}
which is represented diagramaticaly as
\begin{center}
\fcolorbox{white}{white}{
  \scalebox{.8}{
  \begin{picture}(534,149) (67,-53)
    \SetWidth{0.5}
    \SetColor{Black}
    \Text(304,57)[lb]{{\Black{$p+q$}}}
    \SetWidth{1.0}
    \Line[arrow,arrowpos=0.5,arrowlength=8,arrowwidth=3,arrowinset=0.2](82,70)(129,29)
    \GOval(135,11)(18,18)(0){0.882}
    \Line[arrow,arrowpos=0.5,arrowlength=8,arrowwidth=3,arrowinset=0.2](128,-7)(81,-48)
    \Photon(152,11)(205,11){6}{3}
    \Text(233,6)[lb]{{\Black{$=$}}}
    \Line[arrow,arrowpos=0.5,arrowlength=8,arrowwidth=3,arrowinset=0.2](274,70)(333,11)
    \Line[arrow,arrowpos=0.5,arrowlength=8,arrowwidth=3,arrowinset=0.2](332,11)(274,-48)
    \Vertex(332,11){9}
    \Text(368,6)[lb]{{\Black{$+$}}}
    \Line[arrow,arrowpos=0.5,arrowlength=8,arrowwidth=3,arrowinset=0.2](391,70)(432,70)
    \Line[arrow,arrowpos=0.5,arrowlength=8,arrowwidth=3,arrowinset=0.2](432,70)(432,-47)
    \Line[arrow,arrowpos=0.5,arrowlength=8,arrowwidth=3,arrowinset=0.2](432,-47)(392,-47)
    \Line[arrow,arrowpos=0.5,arrowlength=8,arrowwidth=3,arrowinset=0.2](438,-47)(438,70)
    \Line[arrow,arrowpos=0.5,arrowlength=8,arrowwidth=3,arrowinset=0.2](438,70)(502,23)
    \Line[arrow,arrowpos=0.5,arrowlength=8,arrowwidth=3,arrowinset=0.2](502,0)(438,-47)
    \GOval(508,11)(18,18)(0){0.882}
    \Photon(525,11)(578,11){6}{3}
    \Text(67,70)[lb]{{\Black{$\alpha$}}}
    \Text(67,-47)[lb]{{\Black{$\beta$}}}
    \Text(110,57)[lb]{{\Black{$p+q$}}}
    \Text(120,-34)[lb]{{\Black{$p$}}}
    \Text(176,23)[lb]{{\Black{$q$}}}
    \Text(215,6)[lb]{{\Black{$\mu\nu$}}}
    \Text(260,70)[lb]{{\Black{$\alpha$}}}
    \Text(260,-47)[lb]{{\Black{$\beta$}}}
    \Text(310,-34)[lb]{{\Black{$p$}}}
    \Text(380,70)[lb]{{\Black{$\alpha$}}}
    \Text(380,-47)[lb]{{\Black{$\beta$}}}
    \Text(400,80)[lb]{{\Black{$p+q$}}}
    \Text(410,-68)[lb]{{\Black{$p$}}}
    \Text(390,6)[lb]{{\Black{$p-r$}}}
    \Text(473,52)[lb]{{\Black{$r+q$}}}
    \Text(473,-39)[lb]{{\Black{$r$}}}
    \Text(548,23)[lb]{{\Black{$q$}}}
    \Text(566,-7)[lb]{{\Black{$\mu\nu$}}}
  \end{picture}
  }
}
\end{center}
This recursion relation is very similar to the one satisfied by the current vertex, the only difference being in the inhomogeneous term $U_0^{\mu\nu}$, which we draw on the diagram as
a larger black dot in the vertex.
Their form and all other details relevant to the computation of the Kubo formulas 
are collected in appendix \ref{app:stresscorrelator}. At the end of the day we find
\begin{align}
	&\zeta (\omega ) = \frac{N i (c_0 - \lambda \mu)^2}{32 \pi \lambda \omega (\lambda c_0 - \mu) \left( (\omega+ 2 c_0) + ( \omega - 2 c_0 ) e^{2 \lambda \arctanh \frac{\omega}{2\mu} }\right)} \bigg(- ( \omega + 2 c_0 )( \mu (\omega - 2 c_0 ) - \lambda c_0 \omega ) \nonumber \\
	&\qquad \qquad \qquad \qquad \qquad \qquad \qquad + ( \omega - 2 c_0 ) ( \mu ( \omega + 2 c_0 ) - \lambda c_0 \omega ) e^{2 \lambda \arctanh \frac{\omega}{2 \mu} } \bigg), \nonumber \\
	&\eta (\omega ) = \frac{N i}{1536 \pi \lambda \omega^2} \bigg( -24 \omega (c_0 - \lambda \mu ) ( \omega^2 - 4 c_0^2 + 4 \lambda \mu (c_0 - \lambda \mu ) - 8 \lambda^2 \mu^2)-160 \lambda (1-\lambda^2)\mu^3 \omega \nonumber \\
	& \qquad \qquad \qquad - 3 ( \omega - 2 c_0 )^3( \omega + 2 c_0 ) e^{2 \lambda \arctanh \frac {\omega}{2 \mu}}+ 3 ( \omega - 2 c_0 )( \omega + 2 c_0 )^3 e^{-2 \lambda \arctanh \frac {\omega}{2 \mu}} \bigg), \nonumber \\
	&\eta_H ( \omega ) = -\frac{N}{512 \pi \lambda \omega^2} \bigg(-2(\omega^4 + 8 c_0^2 \omega^2 - 16 c_0^4) + 16 (c_0 - \lambda \mu )^2 \omega^2 \nonumber \\
	&\qquad \qquad \qquad \qquad+ (\omega-2 c_0 )^3 ( \omega+ 2 c_0) e^{2 \lambda \arctanh \frac{\omega}{2 \mu} } + ( \omega - 2 c_0 ) (\omega + 2 c_0 )^3 e^{-2 \lambda \arctanh \frac{\omega}{2 \mu} }\bigg) .
\end{align}

\begin{figure}
\begin{center}
\includegraphics[width=.45\textwidth]{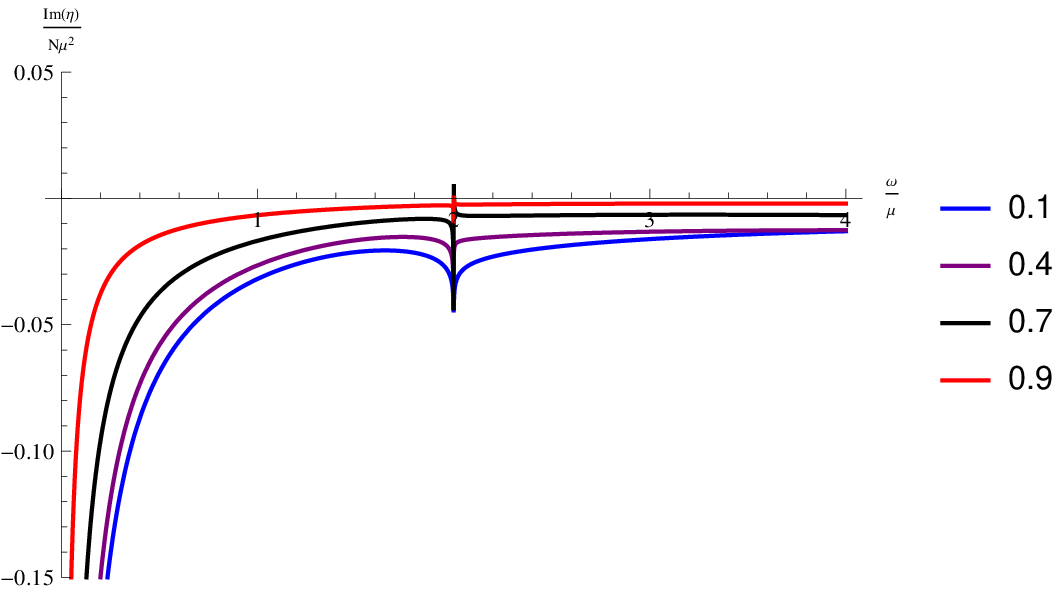}
\includegraphics[width=.45\textwidth]{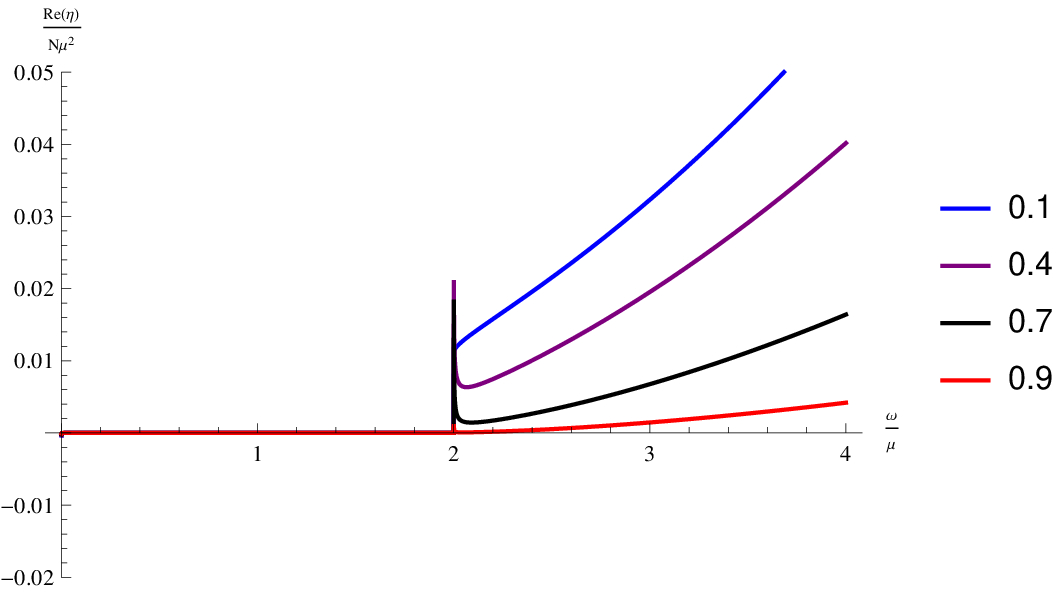}
\end{center}
\caption{Shear viscosity $\eta$ at $m_0 / \mu = .3$ and several values of $\lambda$.}
\label{fig:ShearViscosity}
\end{figure}
\begin{figure}
\begin{center}
\includegraphics[width=.45\textwidth]{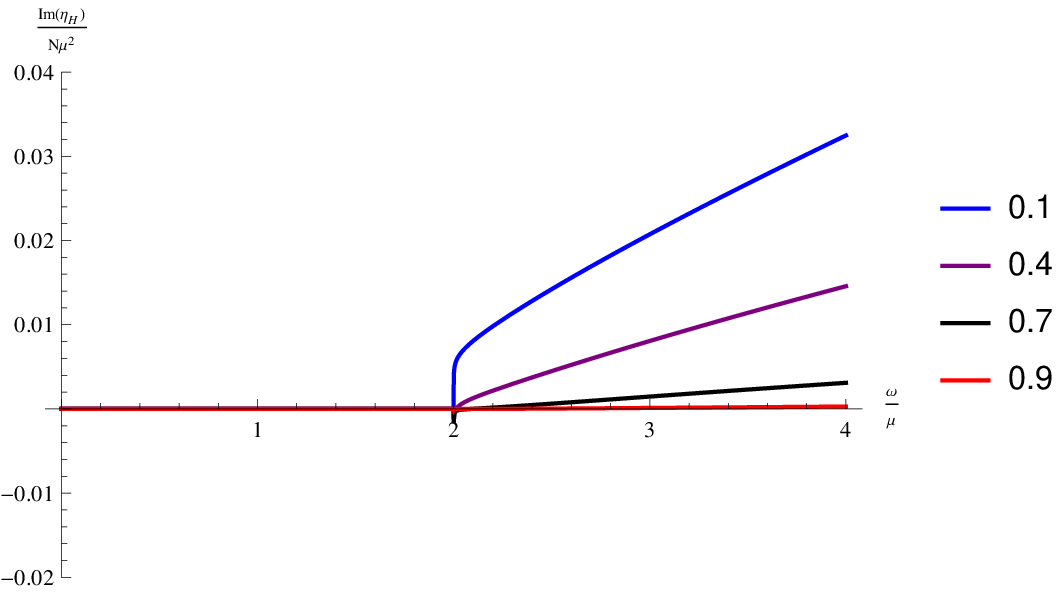}
\includegraphics[width=.45\textwidth]{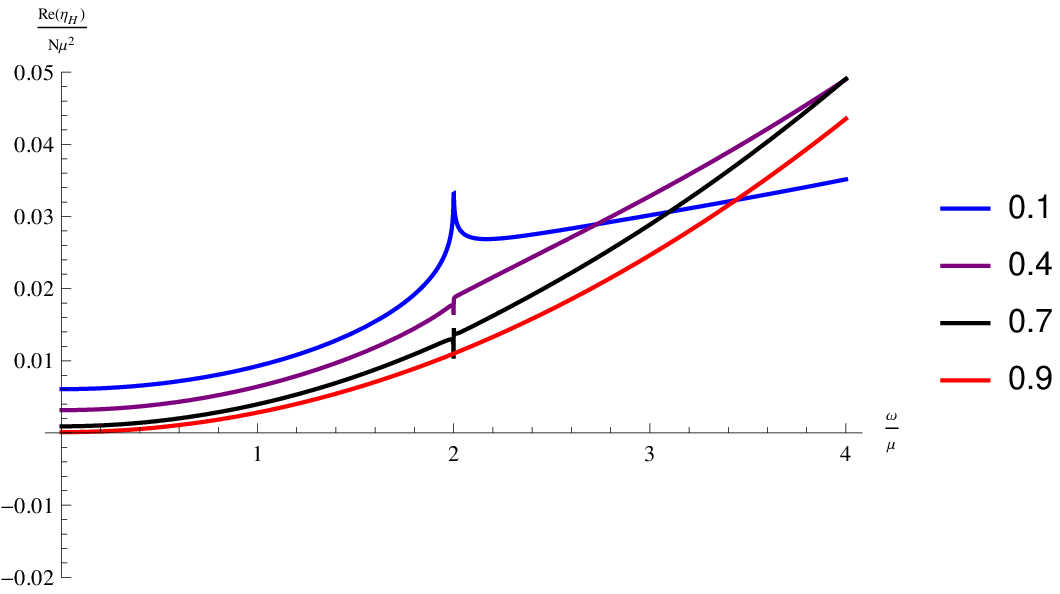}
\end{center}
\caption{Hall viscosity $\eta_H$ at $m_0 / \mu = .7$ and several values of $\lambda$.}
\label{fig:HallViscosity}
\end{figure}
\begin{figure}
\begin{center}
\includegraphics[width=.45\textwidth]{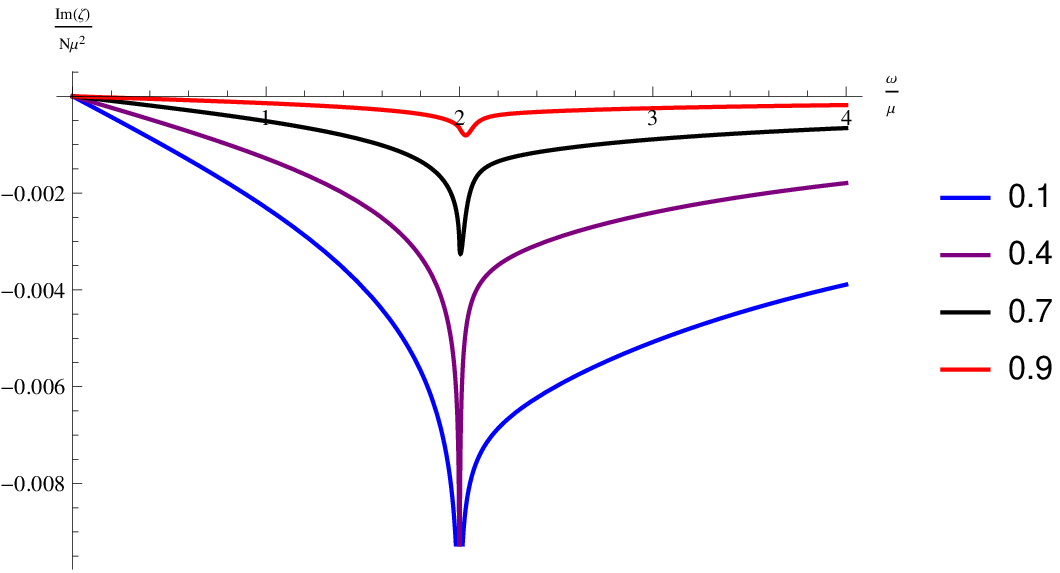}
\includegraphics[width=.45\textwidth]{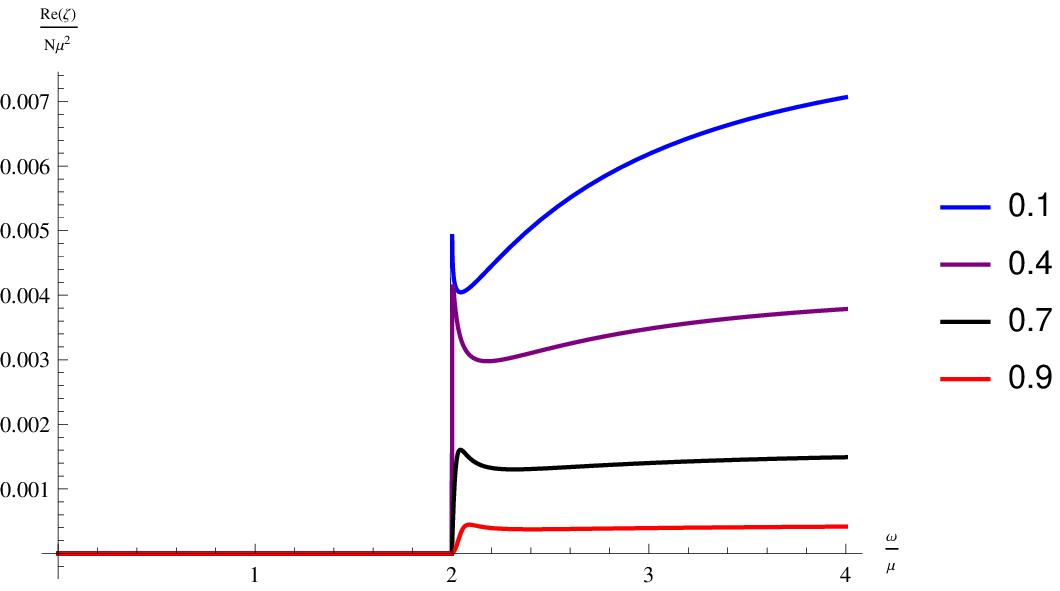}
\end{center}
\caption{Bulk viscosity $\zeta$ at $m_0 / \mu = .8$ and several values of $\lambda$.}
\label{fig:BulkViscosity}
\end{figure}

See figures \ref{fig:ShearViscosity}, \ref{fig:HallViscosity}, \ref{fig:BulkViscosity} for illustration. As with the conductivity, these exhibit a discontinuity as one crosses the pair-production threshold. Imaginary part of the shear viscosity has a pole at zero frequency, which agrees with the LFL theory prediction, stating that $\eta_{DC}\sim \tau$, where quasiparticle life-time, $\tau\sim 1/T^2$, is infinite at zero temperature, see, {\it e.g.}, \cite{baym2008landau}. We also note that unlike the shear viscosity, the bulk viscosity approaches a finite limit at low frequencies (in fact, it goes to zero), a general feature of the bulk viscosity within any quantum field theory, proven by Bradlyn, Goldstein and Read in \cite{Bradlyn2012}.

In the high frequency limit these simplify to,
\begin{align}
	&\zeta (\omega ) = - \frac{N m_0^2}{32 \pi \lambda}  (1 \mp \lambda)^2 \tan \frac{\pi \lambda}{2}, \nonumber \\
	&\eta (\omega ) = - \frac{N \sin \pi \lambda}{256 \pi \lambda}  \omega^2 
	+\frac{i(\lambda-c_0)\sin^2\left(\frac{\pi\lambda}{2}\right)\,\omega}{32\pi\lambda},
	\label{highfrequencyviscosity} \\
	&\eta_H (\omega ) = \frac{N \sin^2 \frac{ \pi \lambda}{2}}{128 \pi \lambda}  \omega^2
	+\frac{i(\lambda-c_0)\sin\, (\pi\lambda)\,}{64\pi\lambda}.
	\nonumber
\end{align}

The nonrelativistic limit $\Delta \mu, \omega \ll | m_0|$ limit of the Hall viscosity is particularly noteworthy. Momentarily restoring SI units, we have
\begin{align}
	&\eta_H (\omega ) = \mp \frac{\hbar}{4}( 1 \mp \lambda ) n + \mathcal O (c^{-2}) ,
\end{align}
which is Read's formula for the Hall viscosity of a non-relativistic fluid of anyons with spin $\frac{1}{2}(1 \mp \lambda )$ \cite{Read2009}. Though this has been proven via adiabatic arguments for non-relativistic gaped states \cite{Read2011} and demonstrated in some examples in \cite{Hoyos:2013eha}.
this is the first known example to our knowledge of a gapless system exhibiting this behavior and suggests the relation may be more general than the current literature indicates.
It would be interesting to investigate precisely how general this relation is.

\section{Discussion}

In this paper we performed an investigation of the large $N$
limit of Chern-Simons theory coupled to a massive fundamental fermions.
Broadly stated, our goal was to analyze this system from a condensed matter physics
point of view, matching it against the phenomenological Landau Fermi liquid framework,
as well as to calculate various thermodynamic and transport observables.

An important question is how well the Chern-Simons-Fermion system agrees
at low temperatures ($T/\mu\ll 1$) with the Landau Fermi liquid theory.
The properties of an LFL system
are encoded in the Landau parameters, which represent
the strength of quasiparticle interaction on the Fermi surface.
The Landau parameters can be calculated microscopically by evaluating quasiparticle scattering amplitudes.
Knowing the Landau parameters,
one can in particular describe a low-temperature thermodynamics of the Fermi liquid.

In this paper we found that this logic is correct for the
large $N$ Chern-Simons-Fermion theory. However, we identified a few
subtleties and interesting properties, not covered by the Landau Fermi-liquid theory, or lying outside of
its regime of applicability.

An explicit calculation of the quasiparticle scattering amplitude
showed that the only non-vanishing Landau parameter of the system is $F_0$.
Assuming LFL theory to be correct, and using
the calculated Landau parameters,
we then found various quantities, such as quasiparticle effective mass,
compressibility and entropy density. On the other hand, all these quantities can be
independently found if one knows the equation of state of the system.
This has already been calculated exactly in the literature for large$N$ Chern-Simons-matter systems. Using these results, we performed some consistency checks for LFL theory.

One notable difference from a standard Landau Fermi liquid appears
in the calculation of entropy density. A general LFL theory implies that the
low-temperature entropy is characterized by the law $s\simeq m^* T$,
which exhibits two important features. The first is a linear dependence
of entropy on temperature. The second is that all the dependence on interaction
strength sits in the quasiparticle effective mass $m^*$,
which in the LFL theory is determined by the Landau parameters.
Derivation of this expression for the entropy relies on the Fermi-Dirac
distribution for the quasiparticles.

We found that in the Chern-Simons-Fermion system the latter assumption is incorrect.
Knowing the fermionic two-point function one can explicitly
calculate the occupation number and in the Chern-Simons-Fermion system it turns
out that the effect of the holonomies of the gauge field
along the thermal circle modifies the quasiparticle distribution function.
Accounting for this, we found that
$s\sim (1-\lambda^2) m^*T$, where $m^*$ is still the LFL quasiparticle effective mass,
and the factor of $1-\lambda^2$ is due to holonomies.

In the non-relativistic limit $T,\mu-|m_0|\ll |m_0|$, where $|m_0|$
is the gap energy, it is technically convenient to study thermodynamic
properties of the system for a wide range of temperatures.
We used this to numerically calculate the temperature dependence
of the heat capacity in the non-relativistic theory. We found an intermediate temperature range, $\frac{n}{N |m_0|}\ll T\ll \frac{n}{N (1-\lambda)|m_0|}$,
between the low-temperature Fermi-liquid state, and the
high-temperature ideal-gas state, which
opens up as the 't~Hooft coupling $\lambda$ approaches 1.
The heat capacity also appears to be linear in this regime, with the slope equal to $(\pi/6)(1-\lambda)$.

Taking large $N$ techniques, we calculated the zero-temperature two-point functions for the $U(1)$ charge
current and the stress tensor and obtained the
the conductivities and viscosities. 
The longitudinal conductivity at low frequencies $\omega/\mu\ll 1$
agrees with the Drude model, with quasiparticle density consistent
with our thermodynamic result (and Luttinger's theorem),
and quasiparticle life-time being infinite.
A zero-frequency pole in the imaginary part of
the shear viscosity also agrees with the zero-temperature LFL prediction.

Our results also allowed us to test
various statements existing in the literature. 
It was pointed out \cite{Haldane:2004zz} that Landau Fermi-liquid
theory is not sufficient to calculate the Hall conductivity, and that the
latter receives an extra contribution from the Berry flux through the Fermi surface.
We have found that the Berry flux expression of \cite{Haldane:2004zz}
does not fully describe Hall conductivity of an interacting Chern-Simons-Fermion system.

As opposed to the shear viscosity,
the bulk viscosity does not diverge
at zero frequency \cite{Bradlyn2012}. We have verified that this general statement
is indeed correct for the Chern-Simons-Fermion system.
One technical subtlety which we encountered in derivation
of the bulk viscosity involves a regularization of the two-point
function for the $T^{+-}$ component of the stress tensor.
We argued that the correct way to regularize expressions
like $\Lambda^n e^{2\lambda\arctanh\frac{\omega}{\Lambda}}$, $n=1,2,3$,
where $\Lambda$ is a UV cutoff, is to first set the exponent to one,
and then remove all the polynomially divergent term.
This issue only arises in calculation of the bulk viscosity.
It would be good to achieve a better understanding of this subtlety.

\acknowledgments
We thank Eun-Gook Moon, Hong Liu and Matt Roberts for many fruitful discussions, and to Ofer Aharony for numerous comments on an earlier version of the manuscript.  This work is supported, in part, by by the US DOE Grant No. DE-FG02-13ER41958, MRSEC Grant No. DMR-1420709 by the NSF, ARO MURI Grant No. 63834-PH-MUR, Oehme Fellowship, and by a Simons Investigator grant from the Simons Foundation.

\appendix

\section{Specific Heat from Statistics}\label{app:CvHolonomies}

In this appendix we investigate the effect of holonomies on the low temperature thermodynamics of our theory. At low temperatures, the entropy density of a fermionic system vanishes linearly with $T$. In a standard $2+1$ dimensional Fermi liquid, the slope is related to the effective mass $m^\star$ as
\begin{align}
	s= \frac{N}{6} \pi m^\star T \,.
\end{align}
This follows directly from the low temperature form of the Fermi-Dirac distribution as may be seen in detail in section 1.1.3 of \cite{baym2008landau}. In this section we carry out the same analysis in the presence of holonomies.

In our case, due to the Chern-Simons gauge field, the electrons do not obey standard Fermi statistics and this needs to be modified. It is easy to see this from a direct evaluation of the occupation number from the Green's function $G^\alpha{}_\beta (p)= \left \langle \psi_\beta (p) \bar \psi^\alpha (-p) \right \rangle$
\begin{align}
	\left \langle n(p_s) \right \rangle = - \frac{1}{\beta} \int d \alpha \rho ( \alpha ) \sum_n \Tr \left( G( \tilde p) \gamma^3 \right) .
\end{align}
Here the sum is over Matsubara frequencies, shifted by the holonomies
\begin{align}
	\tilde p_3 = \frac{2 \pi\left( n+ \frac 1 2 \right) + i \hat \mu - \alpha }{\beta} .
\end{align}
Taking, $\hat \mu \rightarrow \hat \mu + i \alpha$ in (3.17) of \cite{Yokoyama:2012fa} we find equation (\ref{distribution}) which we reproduce here
\begin{align}
	n(p_s) = \frac{N}{2} \int d \alpha \rho ( \alpha ) \left( \tanh \frac 1 2 ( \beta ( E_p +  \mu )+ i \alpha ) - \tanh \frac 1 2 (\beta( E_p - \mu) - i \alpha )\right) \,.
\end{align}

We only need this in the low temperature limit, in which case we have a Fermi-Dirac distribution, modified in the appropriate manner by the holonomies
\begin{align}\label{FDdistribution}
	n (p_s ) = N \int d \alpha \rho ( \alpha )  \frac{1}{1+ e^{\beta ( E_p - \mu ) - i \alpha}} \,.
\end{align}
Now we simply follow the steps of \cite{baym2008landau}. In terms of the occupation number, the entropy density is\footnote{This assumes the states are in one-to-one correspondence with the free Fermi gas. This may fail, in which case Fermi Liquid theory is not expected to hold. However, it is certainly true in our case.}
\begin{align}
	s = - \int \frac{d p_s}{(2 \pi)^2} \left( n(p_s) \ln n (p_s) + (1 - n (p_s )) \ln ( 1 - n ( p_s ))\right).
\end{align}
We evaluate its variation with respect to the temperature at small $T$
\begin{align}\label{deltas}
	\delta s = - \int \frac{d^2 p}{(2 \pi)^2} \delta n (p_s ) \ln \frac{n (p_s)}{1 - n ( p_s )},
\end{align}
where
\begin{align}
	\delta n(p_s) = \frac{\partial n (p)}{\partial E_p} \left( - \frac{E_p - \mu}{T} \delta T + \delta E_p - \delta \mu \right) .
\end{align}
The term $\delta E_p - \delta \mu$ is higher order in $T$ and will be dropped.

The argument of the log in (\ref{deltas}) depends on the holonomies, but the contribution is subleading at low temperatures
\begin{align}
	\ln \frac{n (p_s)}{1- n(p_s)} = \ln \left( \int d \alpha \rho ( \alpha ) e^{- \beta ( E_p - \mu ) + i \alpha }\right) = \ln \left( \frac{\sin \pi \lambda}{\pi \lambda}e^{- \beta ( E_p - \mu )}\right) \approx - \frac{E_p - \mu}{T} .
\end{align}
Plugging this all into $\delta s$ we find that (1.1.37-38) of \cite{baym2008landau} survives, only the distribution function is modified by holonomies 
\begin{align}
	\delta s &= - \int \frac{d^2 p}{(2 \pi)^2} \frac{\partial n (p_s)}{\partial E (p_s)} \left(\frac{E(p_s) - \mu}{T} \right)^2 \delta T \nonumber \\
	&= - \nu ( E_F ) \int d \alpha \rho ( \alpha ) \int_{- \infty}^\infty dx \frac{\partial}{\partial x}\left( \frac{1}{1 + e^{x- i \alpha}}\right)x^2T \,.
\end{align}
Here $\nu (E_F ) = \frac{N m^\star}{2 \pi}$ is the density of states at the Fermi surface.

For ease of integration we shift $x \rightarrow x + i \alpha$ and then deform the contour back to the real axis (there are no poles on the Riemann sphere to get in the way)
\begin{align}
	s = - \nu ( E_F ) \int d \alpha \rho ( \alpha ) \int_{- \infty}^\infty dx \frac{\partial}{\partial x}\left( \frac{1}{1 + e^{x}}\right)(x+ i \alpha)^2T \,.
\end{align}
Evaluating, we find
\begin{align}
	s = \nu ( E_F ) \int d \alpha \rho ( \alpha ) \left(\frac{\pi^2}{3} -\alpha^2 \right) T = \frac{N}{6} \pi ( 1 - \lambda^2 ) m^\star T \,,
\end{align}
which implies (\ref{CvLowTemp}).

\section{Non-Relativistic Thermodynamics}
\label{App:Non-Relativistic Thermodynamics}
In this section we present the details of the various limits performed in section \ref{sec:DegeneracyTemp}. We begin by taking the non-relativistic limit of the gap equations (\ref{gapEquations}). Denote the zero density, zero temperature pole mass by $m_0$. The theory then has a gap $|m_0|$ and we define $\Delta \mu$ to be the location of the chemical potential relative to the gap: $\mu = |m_0|+ \Delta \mu$.
The non-relativistic limit is achieved by taking $T$ and $\Delta \mu$ to be small compared to the gap
\begin{align}
	\tilde T , \Delta \tilde \mu \ll 1, 
	\qquad \qquad \xi = \frac{\Delta \mu}{ T} ~ \text{arbitrary}.
\end{align}
Here and in what follows, the tilde denotes that a quantity is measured in units of the gap energy, for instance, $\tilde T = T/|m_0|$.

In terms of these variables the gap equations read
\begin{align}
	\tilde c_0 &= \tilde m + 2 \lambda \tilde T \mathcal C ,\nonumber \\
	\mathcal C &= \frac 1 2 \int d \alpha \rho (\alpha ) \left( \ln  2 \cosh \frac 1 2 \left( \frac{|\tilde  c_0|+ 1 }{\tilde T} +   \xi + i \alpha   \right) + \ln  2 \cosh \frac 1 2  \left( \frac{|\tilde  c_0|- 1 }{ \tilde T} -  \xi - i \alpha \right) \right),
\end{align}
and the equation of state is
\begin{align}\label{NRFreeEnergy}
	F = \frac{N V_2 | m_0 |^3}{6 \pi} \bigg( &|\tilde c_0|^3 - 2 ( | \tilde c_0|^2 - \tilde m^2) \tilde T \mathcal C + 2 \lambda \tilde  m \tilde T^2 \mathcal C^2 - f_0 \nonumber \\
	&- 3 \tilde T \int d \alpha \rho ( \alpha ) \int^\infty_{|\tilde c_0|} dz z \left( \ln (1 + e^{- \frac{z+1}{\tilde T} - \xi - i \alpha } )+ \ln ( 1 + e^{-\frac{z-1}{\tilde T} + \xi + i \alpha }) \right)\bigg)  
\end{align}
where $z= \tilde T y$ .

If $c_0$ passes through zero for some range of $\xi$, the temperature and chemical potential will be large in comparison to the rest energy of the quasi-particles. In this case the non-relativistic limit is not sensible. When using the results of this section one should keep this in mind.

\subsection{Solving the Gap Equations}
We first work on solving the gap equations perturbatively in $\tilde T$. Expand $\tilde c_0$ about the zero temperature answer $c_0 = m_0$
\begin{align}
	\tilde c_0 = \pm 1 + a_1 \tilde T + a_2 \tilde T^2  \cdots ,
	&&\text{so that}
	&&| \tilde c_0 | = 1 \pm a_1 \tilde T \pm a_2 \tilde T^2 + \cdots.
\end{align}
Here, as in the main text we have fixed $\lambda >0$ while $m_0$ may have either sign. Recall that the upper sign will refer to $m_0>0$ and the lower sign to $m_0<0$.

Plugging this expansion into $\mathcal C$ we find
\begin{align}
	\mathcal C &= \frac{1}{2 \tilde T} + \frac{\xi}{2} + \frac{1}{2 \pi \lambda} \text{Im} ~ \text{Li}_2 \left( - e^{- \xi \pm a_1 - i \pi \lambda }\right) + \mathcal O (\tilde T ) \nonumber.
\end{align}
Feeding this back into $\tilde c_0 = \tilde m + 2 \lambda \tilde T \mathcal C$ we find
\begin{align}
	&\pm 1 + a_1 \tilde T  + \cdots = \tilde m + \lambda + \left( \lambda \xi + \frac{1}{\pi} \text{Im} ~ \text{Li}_2 \left( - e^{- \xi \pm a_1 - i \pi \lambda }\right)  \right) \tilde T  + \cdots ,
\end{align}
which implies 
\begin{align}
	a_1 = \lambda \xi + \frac{1}{\pi} \text{Im} ~ \text{Li}_2 \left( - e^{- \xi \pm a_1 - i \pi \lambda }\right) ,
\end{align}
while the $\tilde O (\tilde T^0)$ equation is trivial.
This is a trancendental equation that determines $a_1$ as a function of $\xi$. A similar analysis at the next order shows that $a_2 = 0$. Redefining $a_1 \rightarrow \lambda f$ for simplicity, we have 
\begin{align}\label{NRGap1}
	&\tilde c_0 = \pm 1 + \lambda f ( \xi) \tilde T + \mathcal O (\tilde T^3), 
	&&\mathcal C = \frac{1}{2 \tilde T} + \frac 1 2 f ( \xi ) + \mathcal O ( \tilde T^2 ) ,
\end{align}
where $f(\xi)$ solves
\begin{align}\label{NRGapEqn}
	f(\xi ) = \xi + \frac{1}{\pi \lambda} \text{Im} ~ \text{Li}_2 \left( - e^{- \xi \pm \lambda f (\xi) - i \pi \lambda }\right) .
\end{align}

\subsection{Equation of State}
Now that we have $|\tilde c_0|$ and $\tilde T \mathcal C$
to second order in $\tilde T$, we may determine the equation of state to the same order. Restoring SI units, we find that terms of higher order are $1/c^2$ suppressed and so are negligible in the non-relativistic limit. 

The $z$ integral in (\ref{NRFreeEnergy}) may be computed exactly. This gives to our order
\begin{align}
	F = \frac{N V_2 | m_0 |^3}{6 \pi} \int d \alpha \rho (\alpha )\bigg( &-|\tilde c_0|^3 - 2 ( | \tilde c_0|^2 - \tilde m^2) \tilde T \mathcal C + 2 \lambda \tilde  m \tilde T^2 \mathcal C^2 - f_0 \nonumber \\
	&- 3 \tilde T^2 |\tilde c_0| \Li_2 \left( - e^{\frac{|\tilde c_0|-1}{\tilde T} - \xi - i \alpha} \right) - 3 \tilde T^2 |\tilde c_0| \Li_2 \left( - e^{\frac{|\tilde c_0|+1}{\tilde T} + \xi+ i \alpha} \right)  \bigg) \,. 
\end{align}
Plugging in the perturbative solution to the gap equations\footnote{The expansion (\ref{polyLogExpansion})
is helpful here for asymptotics.} we find
\begin{align}\label{appendNREOS}
	F = - \frac{N V_2 |m_0|}{12 \pi} T^2 \bigg(  \left( \pi^2 (1 - \lambda^2 ) + 3 ( \xi^2 \pm \lambda f (\xi) ( f(\xi) - 2 \xi) )\right)
		&+ 6 \int d \alpha \rho ( \alpha ) \Li_2 \left( - e^{- \xi + i \alpha \pm \lambda f(\xi)}\right) \bigg)\,,
\end{align}
demonstrating equation (\ref{NREOS}).

\subsection{Low Temperatures}

Here we analyze the low temperature regime $\xi \gg 1$. Beginning with the gap equation, it's easy to see from (\ref{NRGapEqn}) that in this regime $f$ is approximately linear and corrections are exponentially suppressed
\begin{align}
	f(\xi ) = \xi + b_1 e^{- (1 \mp \lambda) \xi} + b_2 e^{- 2(1 \mp \lambda)  \xi} + \cdots .
\end{align}
These corrections do not enter the equation of state to first order in $e^{- ( 1 \mp \lambda) \xi}$. The pressure $p = - F / V$ in this limit is 
\begin{align}\label{NRp}
	p&= \frac{N}{12 \pi} | m_0 | T^2 \bigg( \pi ( 1 - \lambda^2 ) + 3 (1 \mp \lambda)  \xi^2 + \mathcal O(e^{-2 (1 \mp \lambda)  \xi}) \nonumber \\
		&\qquad \qquad \qquad+ 6 \int d \alpha \rho ( \alpha ) \Li_2 \left( - e^{- (1 \mp \lambda)  \xi + i \alpha } + \mathcal O(e^{-2 (1 \mp \lambda)  \xi})  \right)\bigg) \nonumber \\
		&= \frac{N}{12 \pi} | m_0 | T^2 \bigg( \pi ( 1 - \lambda^2 ) + 3 (1 \mp \lambda)  \xi^2 - 6 \frac{\sin \pi \lambda}{\pi \lambda}e^{- (1 \mp \lambda)  \xi}+ \mathcal O(e^{-2 (1 \mp \lambda)  \xi})\bigg) .
\end{align}

 In section (\ref{sec:DegeneracyTemp}) we require the pressure as a function of temperature and density. For this we need
 \begin{align}
	n = \left( \frac{\partial p}{\partial \Delta \mu}\right)_T= \frac{N}{2 \pi} | m_0 | (1 \mp \lambda) T \left( \xi + \frac{\sin \pi \lambda}{\pi \lambda} e^{- (1 \mp \lambda)  \xi} + \mathcal O \left( e^{- 2 (1 \mp \lambda)  \xi }\right)\right) .
\end{align}
Inverting for $\xi$ we find
\begin{align}
	\xi &= \frac{2 \pi}{1 \mp \lambda } \frac{n}{N | m_0 | T} - \frac{\sin \pi \lambda}{\pi \lambda} e^{ - 2 \pi  \frac{n}{N | m_0| T}} + \mathcal O \left(  e^{ - 4 \pi  \frac{n}{N | m_0| T}} \right) .
\end{align}
Bringing this all together, we find the equation of state as a function of temperature and density
\begin{align}
	\frac{12 \pi}{N | m_0 | T^2}p &= \pi^2 ( 1 - \lambda^2 ) + \frac{12 \pi^2}{| \tilde m | } \left( \frac{n}{N | m_0 | T} \right)^2 \nonumber \\
	&\qquad \qquad - 12 \pi \frac{\sin \pi \lambda}{\pi \lambda}  \frac{n}{N | m_0 | T}  e^{ - 2 \pi  \frac{n}{N | m_0| T}} + \mathcal O \left(  e^{ - 4 \pi  \frac{n}{N | m_0| T}} \right).
\end{align}
The first two terms give a linear specific heat of slope $\frac \pi 6 N ( 1 - \lambda^2) |m_0|$. Corrections to this behavior then  begin at $ \mathcal O \left(  e^{ - 2 \pi  \frac{n}{N | m_0| T}} \right)$ and are numerically small when
\begin{align}
	2 \pi \frac{n}{N | m_0 | T} \gg 1,
	&&\text{i.e.}
	&& T \ll T_q,
	&&\text{where}
	&& T_q = 2 \pi \frac{n}{N |m_0 |},
 \end{align} 
 independent of coupling.

\subsection{Virial Expansion of the Non-Relativistic Equation of State}
In this section we provide the details of the virial expansion used in section \ref{sec:DegeneracyTemp} to investigate the classical limit. This is an expansion in small fugacity $z = e^\xi$, the opposing limit to the one considered above. Here we have
\begin{align}
	f(\xi) = a_0 + a_1 e^\xi  + \cdots ,
\end{align}
which, after plugging in to the gap equation gives
\begin{align}
	a_0 &+ a_1 e^\xi + a_2 e^{2 \xi} + \cdots 
	= \xi + \frac{1}{\pi \lambda} \text{Im} ~ \Li_2 \left( - e^{- \xi \pm \lambda a_0 - i \pi \lambda } \left( 1 \pm \lambda a_1 e^\xi + \left( \pm \lambda a_2 + \frac 1 2 \lambda^2 a_1^2 \right) e^{2 \xi} + \cdots \right) \right) . \nonumber
\end{align}
Using the expansion
\begin{align}\label{polyLogExpansion}
	\Li_n (z) &= (-1)^{n-1} \sum_{k=1}^\infty \frac{1}{k^n z^k} - \frac{(2 \pi i)^n}{n!} B_n \left( \ln (- z) + \frac 1 2\right) ,
\end{align}
 we find that to lowest order
\begin{align}
	a_0 &= \xi + \frac{1}{\pi \lambda} \text{Im} \left( \xi (\pm \lambda a_0 - i \pi \lambda) - \frac{1}{2} ( \pm \lambda a_0 - i \pi \lambda )^2 \right) \nonumber \\
		&= - \frac{1}{2 \pi \lambda} \text{Im} ( \pm \lambda a_0 - i \pi \lambda)^2 = \pm \lambda a_0 \nonumber \\
	\implies \qquad a_0 &= 0.
\end{align}
While to first order (\ref{polyLogExpansion}) gives
\begin{align}
	a_1 = \frac{1}{\pi \lambda} \text{Im} \left( e^{i \pi \lambda} \pm \lambda a_1 ( \xi + i \pi \lambda ) \right)
	&&\implies
	&&a_1 = \frac{1}{1 \mp \lambda } \frac{\sin \pi \lambda}{\pi \lambda} .
\end{align}

Plugging this back into (\ref{appendNREOS}) we find the equation of state is
\begin{align}
	p = \frac{N}{2 \pi}  \frac{\sin \pi \lambda}{\pi \lambda}  |m_0| T^2 e^\xi \left(  1- \frac{2 \sin \pi \lambda + \pi (\pm 1 - \lambda) \cos \pi \lambda}{4 \pi (\pm 1 - \lambda)}e^{\xi} \right) + \mathcal O (e^{3 \xi} ) .
\end{align}
We need this as a function of temperature and density. 
The density is
\begin{align}
	n = \frac{N }{2 \pi}  \frac{\sin \pi \lambda}{\pi \lambda} |m_0| T e^\xi \left(  1 - \frac{2 \sin \pi \lambda + \pi \tilde m \cos \pi \lambda}{2 \pi \tilde m}e^{\xi} \right) + \mathcal O (e^{3 \xi} ) .
\end{align}
Inverting this and we can rearrange the equation of state into a virial
expansion
\begin{align}
	\frac{p}{n T} = 1 + v_2 \frac{n}{N |m_0| T} + \cdots ,
	\qquad \text{where} \qquad
	v_2 =  \frac{\pi \lambda}{\pm 1-\lambda}+ \frac{1}{2} \pi^2 \lambda \cot  \pi \lambda .
\end{align}
$v_2$ is the second virial coefficient and determines the size of deviations from the ideal gas law to lowest order in the classical limit.

\section{Details of the Four-point Vertex Calculation}\label{app:4ptVertexCalc}
In this appendix we present the calculation of the four point vertex function (\ref{4ptVertex}). This calculation proceeds essentially along the lines of appendix F of \cite{Jain:2014nza}, with the presence of a Fermi surface being the only new feature. Although this adds an extra layer of complication, the problem is simpler insofar as we only require the answer at the Fermi surface.

As explained in seciton \ref{sec:4ptVertex}, $V$ satisfies a Schwinger-Dyson equation (\ref{RecursionRelation}). Since the gluon propagator has only ``$+3$'' components, this reads
\begin{align}\label{appIntegralEqn}
	V ( p , k , q ) &= - \frac{1}{2} ( I \otimes \gamma^+ - \gamma^+ \otimes I ) G_{+3}(p-k) \nonumber \\
	&- \frac{1}{2} \int \frac{d^3 r}{( 2 \pi )^3} G_{+3} ( r - p ) H_+ \left( G(r+q) V(r,k,q) G(r) \right) ,
\end{align}
where we have used the identity
\begin{align}
	(\gamma^+)^\alpha{}_\beta (\gamma^3)^\gamma{}_\delta - ( \gamma^3 )^\alpha{}_\beta (\gamma^+ )^\gamma{}_\delta = \delta^\alpha{}_\delta (\gamma^+)^\gamma{}_\beta - ( \gamma^+)^\alpha{}_\delta \delta^\gamma{}_\beta.
\end{align}
$H_+$ denotes the operator on matrices
\begin{align}
\label{HplusofA}
	H_+ (A) = \gamma^3 A \gamma^+ - \gamma^+ A \gamma^3 .
\end{align}
Expanding $A$ in the basis $A = A_I I + A_+ \gamma^+ + A_- \gamma^- + A_3 \gamma^3$,
this acts as
\begin{align}
\label{HplusofAexpansion}
	H_+(A) = 2 A_I \gamma^+ - 2A_- I .
\end{align}
Hence the Schwinger-Dyson equation implies an expansion in the product basis of the form
\begin{align}
	V (p,k,q) = g (p,k,q) I \otimes I + g_1 (p,k,q) I \otimes \gamma^+ + f (p,k,q) \gamma^+ \otimes I + f_1 (p,k,q) \gamma^+ \otimes \gamma^+ .
\end{align}

Plugging this into (\ref{appIntegralEqn}), we find the following integral equations for $f$, $g$, $f_1$, $g_1$
\begin{align}\label{gfIntEqn}
	g ( p , k , q_3 ) &= - 4 \pi i \lambda \int \frac{d^2 r}{(2 \pi)^2} \frac{r_-}{(r-p)_-} \frac{\theta (r_s - p_F)}{ 4 E_r^3 } \nonumber \\
	&\left( 2 f ( r,k,q_3) r_- + g (r ,k,q_3)( 2 i \Sigma_I (r) - q_3)\right) , \nonumber \\
	g_1 ( p , k , q_3 ) &= - \frac{2 \pi i \lambda}{(p-k)_-} -  4 \pi i \lambda \int \frac{d^2 r}{(2 \pi)^2} \frac{r_-}{(r-p)_-}\frac{\theta (r_s - p_F)}{ 4 E_r^3 } \nonumber \\
	&\left( 2 f_1 ( r,k,q_3) r_- + g_1 (r ,k,q_3)( 2 i \Sigma_I (r)) - q_3\right) ,  \\
	f ( p , k , q_3 ) &= \frac{2 \pi i \lambda}{ (p - k )_- } -  4 \pi i \lambda \int \frac{d^2 r}{(2 \pi)^2} \frac{1}{(r-p)_-}\frac{\theta (r_s - p_F)}{ 4 E_r^3 } \nonumber \\
	&\left(- f ( r,k,q_3) r_- ( 2 i \Sigma_I (r) + q_3 )+2 g (r ,k,q_3)( \Sigma^2_I(r) - E^2_r   )\right), \nonumber \\
	f_1 ( p , k , q_3 ) &=  -4 \pi i \lambda \int \frac{d^2 r}{(2 \pi)^2} \frac{1}{(r-p)_-}\frac{\theta (r_s - p_F)}{ 4 E_r^3 } \nonumber \\
	&\left( - f_1 ( r,k,q_3) r_- ( 2 i \Sigma_I (r) + q_3 )+2 g_1 (r ,k,q_3)( \Sigma^2_I(r) - E_r^2)\right) .\nonumber
\end{align}
where $E_r = \sqrt{r_s^2 + c_0^2}$ and we have already evaluated the $r^3$ integral.

In the above we have taken $q_{\pm} = 0$ followed by $q_3=0$, in accordance with the order of limits needed in (\ref{InteractionFunction}). As is well known in LFL theory, the double pole singularity in the product $G(r+q) G(r)$ as $q \rightarrow 0$ depends essentially on the order in which this limit is taken (see for instance section 18 of \cite{abrikosov1975methods}).
For now we work in Minkowski signature. In the ``rapid" limit we are considering, a singular term
\begin{align}
	\frac{2 \pi i Z^2 \hat{\mathbf r} \cdot \mathbf q}{q^0 - v_F \hat{\mathbf r} \cdot \mathbf q} \delta (r^0 - \mu) \delta (r_s - p_F)
\end{align}
peaked at the Fermi surface drops out and we are left with only the poles
\begin{align}
	&\frac{1}{(r^0 + E_{ r} - i \epsilon)( r^0 + q^0 + E_{ r} - i \epsilon)}\nonumber \\
	&\qquad \times \frac{1}{(r^0 - E_{ r} + i \epsilon\, \text{sgn}(r_s - p_F ))( r^0+ q^0- E_{ r } + i \epsilon \,\text{sgn}(r_s - p_F ))}
\end{align}
in the product of two propagators.

The placement of $i \epsilon$'s in this equation is essential: we have $+ i \epsilon$ for poles above the Fermi surface and $- i \epsilon$ for those below. This is simply a generalization of the Feynman prescription in the presence of a Fermi surface and can be confirmed to be the correct prescription in the same way. We then have that when $r_s < p_F$ all poles lie above the real axis. The contour may then be closed below and the $r^0$ integration yields zero. When $r_s > p_F$ the two poles on the left hand side are above the axis while the two on the right hand side are below. We may then Wick rotate to Euclidean space and perform the $r^3$ integrals to obtain (\ref{gfIntEqn}).
This is why the integrals over spatial momentum are restricted to be above the Fermi surface.
At this point we may safely take $q^3 \rightarrow 0$.

The equations (\ref{gfIntEqn}) are a bit of a mess, but are luckily very similar in form to those of \cite{Jain:2014nza}. In particular the angular dependence is identical.
We then use the same ansatz present in their work,
\begin{align}
	g(p,k) &= - \frac{1}{2} \frac{p_-}{(p-k)_-} W_0 (x,y) + \frac{1}{2} W_1 (x,y) , \nonumber \\
	f (p,k) &= \frac{1}{2} \frac{1}{(p-k)_-} W_3 (x,y) - \frac{p_+}{p_s^2} W_2 (x,y) , \nonumber \\
	g_1 (p,k) &= \frac{1}{2} \frac{k_+ p_-}{(p-k)_-} B_2 (x,y) + \frac{1}{2} \frac{1}{(p-k)_-} B_3 (x,y) , \nonumber \\
	f_1 (p,k) &= - \frac{1}{p_s^2} \frac{p_+}{(p-k)_-} B_0 (x,y) - \frac 1 2 \frac{k_+}{(p-k)_-} B_1 (x,y) .
\end{align}
where $x = 2 E'_p$ and $y=2E'_k$.This completely fixes the angular dependence of the solution.
Plugging this in and performing the angular integrals we obtain a system of ordinary integral equations
\begin{align}
	&W_0 = - \frac{i \lambda}{\mu} \int_y^x d x' \frac{X W_0  + 2 W_3}{x'^2}  \label{W0 Int}, \\
	&W_1 = - \frac{i \lambda}{\mu} \int_y^\infty d x' \frac{X W_0 + 2 W_3 }{x'^2}
+  \frac{i \lambda}{\mu} \int_x^\infty d x' \frac{X W_1 + 2 W_2 }{x'^2} \label{W1 Int}, \\
	&W_2 = - \frac{i \lambda}{\mu} \int_{2}^x d x' \frac{Y_1 W_1 + Y W_2 }{x'^2}, \label{W2 Int} \\
	&W_3 = 4 \pi i \lambda + \frac{i \lambda}{\mu} \int_x^y d x' \frac{Y_1 W_0 + Y W_3 }{ x'^2},  \label{W3 Int}
\end{align}
\begin{align}
	&B_0 = - \frac{i \lambda}{\mu} \int_{2 }^x d x' \frac{Y B_0 + Y_1 B_3}{x'^2} \label{B0 Int} , \\
	&B_1 = - \frac{8 i \lambda}{\mu^3} \frac{1}{4 \lambda^2 - y^2} \int_{2 }^y d x' \frac{Y B_0 + Y_1 B_3}{x'^2}- \frac{i \lambda}{\mu} \int_y^x d x' \frac{Y B_1 + Y_1 B_2}{x'^2} , \label{B1 Int}  \\
	&B_2 =  \frac{i \lambda}{\mu} \int_x^\infty d x' \frac{2 B_1 + X B_2}{x'^2}  \label{B2 Int} ,\\
	&B_3 =  -4 \pi i \lambda + \frac{i \lambda}{\mu} \int_x^y d x' \frac{2 B_0 + X B_3}{x'^2}- \frac{i \lambda \mu}{8} (y^2 - 4 \lambda^2 ) \int_y^\infty \frac{2 B_1 + X B_2 }{x'^2} , \label{B3 Int}
\end{align}
where we have denoted
\begin{align}
\label{Y1xydefinitions}
	& X = q_3 - 2 i \Sigma_I ,
	&& Y = q_3 + 2 i \Sigma_I,
	&&Y_1=2\Sigma_I^2-\frac{\mu^2}{2} x^2\,.
\end{align}
Of course, all integrals are understood to terminate once one of the limits dips below the Fermi surface at $x'=2$.
For $x>2$ we have
\begin{align}
	&X(x)=-2im-i\lambda\mu x,
	\qquad \qquad Y(x)=2im + i\lambda\mu x, \nonumber \\
	&Y_1(x)=\frac{1}{2}\mu^2x^2(\lambda^2-1)+2m(\lambda\mu x+m) .
\end{align}

If we seek only the solution at the Fermi surface, some of the $W$'s and $B$'s may be simply read off from the above equations
\begin{align}
	W_0 = W_2 = 0 , \qquad W_3 = 4 \pi i \lambda, \qquad B_0 = B_1 = 0, \qquad \text{at } x = y = 2 .
\end{align}
Unfortunately, the solutions for $W_1, B_2$ and $B_3$ require knowledge of all functions in our decomposition away from the Fermi surface and we are forced to solve all the equations for arbitrary $x>2$, $y>2$.
These equations are not difficult to solve.
To proceed, first differentiate the integral equations with respect to $x$ to obtain ordinary differential equations. Solving the differential equations produces
constants of integration which are functions of $y$.
These are fixed by plugging the solution back in to the integral equations and demanding consistency.
All these steps are straightforward to perform in {\it Mathematica}.



After the dust settles we obtain
\begin{align}
	W_0&=0\,,\quad W_1=\frac{4\pi\lambda^2}{\mu -\lambda c_0}\,,\quad W_2=0\,,\quad W_3=4\pi i\lambda ,\\
	B_0&=0\,,\quad B_1=0\,,\quad B_2=0\,,\quad B_3=-4\pi i \lambda\,,
\end{align}
which gives the result (\ref{4ptVertex}).

\section{Details of the Current Vertex Calculation}
\label{app:currentvertex}

In this appendix we provide details for the calculation of the current-current correlation function.
Let us begin by solving the recursion relation (\ref{vertexrecursionrelation})
for the current vertex. Using (\ref{HplusofA})
this reads
\begin{align}
	V^\mu( p ; q ) &=i\gamma^\mu
	- 2 \pi i \lambda \int \frac{d^3 r}{( 2 \pi )^3} \frac{1}{(r-p)_-} H_+ \left( G(r+q) V^\mu(r;q) G(r) \right) ,
\end{align}
where we have dropped the spin indices.
The three-point vertex function then must take the form
\begin{align}
	&V^+  (p;q) = \gamma^+ F_1(x,z) - I\, p_-'\,G_1(x,z) ,\\
	&V^- (p;q) = i\gamma^- + \gamma^+\, 2\,\frac{p_+^{\prime 2}}{p_s^{\prime 4}}\,F_2(x,z) - I\, \frac{p_+'}{p_s^{\prime 2}}\,G_2(x,z)   .
\end{align}
where we denoted $z=q_3'$, and $x= 2 E'_p$ is as before.

Plugging these into the recursion relation, taking $q_\pm = 0$, and performing the $r_3$ integration we find
\begin{align}
	&G_1 = i \lambda \int_x^{2 \Lambda'} d x' \frac{X G_1  + 2 F_1}{z^2 + (x')^2} , \label{g Int} \\
	&F_1 = i + i \lambda \int_x^{2 \Lambda'} d x' \frac{Y_1 G_1 + Y F_1 }{z^2 + ( x' )^2} , \label{f Int}  \\
	G_2 &=-i \lambda \int_2^x d x' \frac{U_2 + X G_2 + 2 F_2}{z^2 + (x')^2} , \\
	F_2 &=- i \lambda \int_2^x d x' \frac{ V_2 + Y_1 G_2 + Y F_2}{z^2 + ( x')^2} ,
\end{align}
where 
\begin{align}
	& X = z - 2 i \Sigma'_I ,
	&& Y = z + 2 i \Sigma'_I,
	&&Y_1=2\Sigma'^2_I-\frac{1}{2} x^2\, \nonumber \\
	&U_2 = \frac i 2 X Y,
	&&V_2 = - I \left( 2 \Sigma'_I ( \Sigma'_I + i z ) + \frac 1 2 x^2 \right) ,
\end{align}
and we introduced cutoff $\Lambda'$ on the spatial momentum $p_s'$.

These equations can be solved along the same lines as the vertex function, giving
\begin{align}
	F_1 &= \frac{2 m' +  \lambda x +i z - (2  m' +  \lambda x -iz) e^{-2 i \lambda \left( \arctan \frac x z - \arctan \frac {2 \Lambda'} z \right)}}{2 z }	 ,\\
	G_1 &= i\,\frac{e^{-2 i \lambda \left( \arctan  \frac x z - \arctan \frac {2 \Lambda'} z \right) }-1}{z },\\
	F_2 &= \frac{1}{8 z} (2 \lambda +2 m+i z)  \bigg( e^{2 i \lambda \left( \arctan
   \left(\frac{z}{2}\right) - \arctan
   \left(\frac{z}{2}\right) \right)}(2 \lambda
   +2 m-i z) (2 m+\lambda  x+i
   z),\nonumber \\
   	&\qquad \qquad -(2 \lambda +2 m+i y) (2 m+\lambda  x-i z) \bigg) \\
	G_2 &= \frac{1}{4 z}\bigg( i \left(4 \lambda ^2+4 m^2+8 \lambda  m+4 i m
   z+2 i \lambda  x z-z^2\right) ,\nonumber \\
   &\qquad \qquad-i (2 \lambda +2 m+i
   z)^2 e^{2 i \lambda  \left(\arctan
   \left(\frac{z}{x}\right)-\arctan
   \left(\frac{z}{2}\right)\right)} \bigg).
\end{align}
In the conformal limit and at vanishing chemical potential our result reproduces
that of \cite{GurAri:2012is}.

\section{Details of the Viscosity Calculation}
\label{app:stresscorrelator}

\subsection{Two Point Correlators}
The viscosity Kubo formula requires both the two point function $\left \langle T^{ij} (q) T^{kl} (-q)\right \rangle$ and the contact term $\left \langle \frac{\delta T^{ij}}{\delta g^{kl}}\right \rangle (q)$. The stress-stress two point function follows exactly along the lines of the current-current two point function, described in appendix \ref{app:currentvertex}, and so we begin there.
The stress tensor vertex satisfies the recursion relation
\begin{align}
	U^{\mu\nu}( p ; q ) &=U_0^{\mu\nu}( p ; q )
	- 2 \pi i \lambda \int \frac{d^3 r}{( 2 \pi )^3} \frac{1}{(r-p)_-} H_+ \left( G(r+q) U^{\mu\nu}(r;q) G(r) \right) .
\end{align}
The inhomogeneous term $U_0^{\mu\nu}( p ; q )$ in the recursion relation
is given by
\begin{center}
\fcolorbox{white}{white}{
  \scalebox{.5}{
  \begin{picture}(660,226) (79,-79)
    \SetWidth{1.0}
    \SetColor{Black}
    \Vertex(192,34){15}
    \Line[arrow,arrowpos=0.5,arrowlength=12,arrowwidth=5,arrowinset=0.3](80,146)(192,34)
    \Line[arrow,arrowpos=0.5,arrowlength=12,arrowwidth=5,arrowinset=0.2](192,34)(80,-78)
    \Line[arrow,arrowpos=0.5,arrowlength=12,arrowwidth=5,arrowinset=0.2](192,34)(80,-78)
    \Vertex(368,34){7.071}
    \Line[arrow,arrowpos=0.5,arrowlength=12,arrowwidth=5,arrowinset=0.2](256,146)(368,34)
    \Line[arrow,arrowpos=0.5,arrowlength=12,arrowwidth=5,arrowinset=0.2](368,34)(256,-78)
    \Line[arrow,arrowpos=0.5,arrowlength=12,arrowwidth=5,arrowinset=0.2](432,146)(544,34)
    \Line[arrow,arrowpos=0.5,arrowlength=12,arrowwidth=5,arrowinset=0.2](544,34)(432,-78)
    \Line[arrow,arrowpos=0.5,arrowlength=12,arrowwidth=5,arrowinset=0.2](736,34)(624,-78)
    \Line[arrow,arrowpos=0.5,arrowlength=12,arrowwidth=5,arrowinset=0.2](624,146)(736,34)
    \PhotonArc(520,10)(33.941,45,225){7.5}{5.5}
    \PhotonArc[clock](712,58)(33.941,-45,-225){7.5}{5.5}
    \Text(256,34)[lb]{\Large{\Black{$=$}}}
    \Text(432,34)[lb]{\Large{\Black{$+$}}}
    \Text(608,34)[lb]{\Large{\Black{$+$}}}
  \end{picture}
}
}
\end{center}
The first term in the r.h.s. originates from the bi-fermionic
part of the stress tensor
\begin{align}
	T^{\mu \nu} = - \frac 1 2 \bar \psi \gamma^{(\mu} \overset{\leftrightarrow} D{}^{\nu)} \psi+ \left( \frac{1}{2} \bar \psi \gamma^\lambda \overset{\leftrightarrow} D_\lambda \psi + m \bar \psi \psi + p_0\right) \eta^{\mu \nu} ,
\end{align}
while the other two terms
come from the $\bar\psi A \psi$ part of the stress tensor
(the internal fermionic lines are full propagators).
In the loop we have the full fermionic propagator. We find
\begin{align}
	&U_0^{++} ( p ; q ) =  - i p_- \gamma^+ ,\nonumber \\
	&U_0^{--} ( p ; q ) = - i \mu \frac{p'_+}{p'^2_s} V_2 \gamma^- - \mu \frac{ p^{\prime 2}_+}{p^{\prime 4}_s} U_2 I ,\nonumber \\
	&U_0^{+-} ( p ; q ) = 2 \mu \frac{p'_+}{p'^2_s}  V_3 \gamma^+ + \frac i 2  p_- \gamma^- + \frac i 2 ( 2 p_3 + q_3 ) \gamma^3 - \mu U_3 I,
\end{align}
where
\begin{align}
	&U_2 = \frac{ \lambda}{12 } \left( 12 m'^2 ( x - 2 ) + 6 \lambda m' ( x^2 - 4 ) + ( \lambda^2 -1 ) (x^3 - 8 )  \right) ,\nonumber \\
	& V_2 =  \frac 1 4 x^2 - c'^2_0 + \frac {\lambda}{4 } (x-2)(4 m' + ( 2 + x ) \lambda ) ,\nonumber \\
	&U_3 = -  m' + \frac 3 4 \lambda  ( 2 \Lambda' - x ) , \nonumber \\
	&V_3 = \frac i 4   \left(  \frac 1 4 x^2 - c'^2_0 - \frac{3 \lambda}{4} ( x - 2 ) ( 4 m' + ( 2 + x ) \lambda )\right) .
\end{align}
We use the ansatz for the matrix structure
\begin{align}
	&V^{++} = \gamma^+ \,F_1(x,z)p_- -  I\, G_1(x,z)\frac{p_-^2}{\mu} ,\nonumber \\
	&V^{--} = - i \mu \frac{p'_+}{p'^2_s} V_2  \gamma^- +  \gamma^+\,
	\frac{2\mu p_+^{\prime 3}F_2(x,z)}{p_s^{\prime 3}}  -I\,
	\frac{\mu p_+^{\prime 2}G_2(x,z)}{p_s^{\prime 4}} , \nonumber \\
	&V^{+-} = \frac i 2 p_-\gamma^- + \frac i 2 ( 2 p_3 + q_3 ) \gamma^3 + \gamma^+\,
	\frac{2\mu p_+' F_3(x,z)}{p_s^{\prime 2}} -  I\,\mu G_3(x,z) ,
\end{align}
which gives the following integral equations
\begin{align}
	G_1 &= i \lambda \int_x^{2 \Lambda'}  dx' \frac{2 F_1 + X G_1}{z^2+ x'^2},\nonumber\\
	F_1 &= - i + i \lambda \int_x^{2 \Lambda'}  dx' \frac{Y F_1 + Y_1 G_1}{z^2 + x'^2},\nonumber\\
	G_2 &= U_2 - i \lambda \int_2^x d x' \frac{1}{y^2 + x'^2} \left( J_2 + 2 F_2 + X G_2 \right) ,
	\nonumber \\
	F_2 &= - i \lambda \int_2^x d x' \frac{1}{z^2 + x'^2} \left( I_2 + Y F_2 + Y_1 G_2 \right),
	\nonumber\\
	G_3 &= U_3 + i \lambda \int_x^{2 \Lambda'} d x' \frac{1}{z^2 + x'^2} \left( J_3 + 2 F_3 + X G_3 \right),\nonumber \\
	F_3 &= V_3 - i \lambda \int_2^x d x' \frac{1}{z^2 + x'^2} \left( I_3 + Y_1 G_3 + Y F_3 \right),
	\nonumber 
\end{align}
where
\begin{align}
	 I_2 &= - \frac i 2 X Y_1 V_2 , \nonumber \\
	J_2 &= i \left( 2 \Sigma'_I ( \Sigma'_I + i z ) + \frac 1 2 x'^2 \right) V_2  , \nonumber \\
	I_3 &= - \frac 1 4 \left( \frac i 4 z x^2 + 2 \left( \frac 5 4 x^2 + z^2 \right) \Sigma'_I - i z \Sigma'^2_I - 2 \Sigma'^3_I \right),\nonumber \\
	J_3 &= \frac i 2 \left( z^2 + \frac 3 4 x^2 - \Sigma'_I ( \Sigma'_I + i z )\right) .
\end{align}

Solving these equations we obtain
\begin{align}
	G_1 &= \frac{e^{-2 i \lambda \left( \arctan \frac x z - \arctan \frac {2\Lambda'} z \right) } - 1}{i z},\nonumber \\
	F_1 &= \frac{(z - 2 i m' - i \lambda x )+(z + 2 i m' + i \lambda x) e^{-2 i \lambda \left( \arctan \frac x z - \arctan \frac {2\Lambda'} z \right) }}{2 i z}, \nonumber\\
	G_2 &=\frac{1}{24z}\left(6 \lambda ^2 \left(16 i m^{\prime 2}+8 m' ((x-2) x-2) z
	+i \left(x^2-4\right) z^2\right)\right.\nonumber\\
	&-\left.16 \lambda  \left(z \left(\lambda ^2+3 m^{\prime 2}
	+3 \lambda  m'-1\right)-3 i \lambda  (\lambda +2 m')^2\right)\right.\nonumber\\
	&+\left.
	2 \lambda  \left(96 i m^{\prime 3}+24m^{\prime 2} (x-4) z+12 i m'(x-2) z^2
	-x z \left(x^2+3 z^2\right)\right)\right.\nonumber\\
	&+\left.3 (2 \lambda +2 m'-i z) (2 i \lambda +2 i m'-z)^3
	 e^{-2 i \lambda  \left( \arctan\left(\frac{z}{2}\right)- \arctan\left(\frac{z}{x}\right)\right)}\right. 
	 \nonumber\\
	 &+\left.3 (z+2 im') (2 m'+i z)^3+8 \lambda ^3 x \left(x^2-6\right) z\right),\nonumber\\
F_2&=\frac{1}{48 z}\left(3 (2 \lambda +2 m'-i z) (2 \lambda +2 m'+i z)^3 (2m'+\lambda  x-i z)
e^{-2 i \lambda  \left( \arctan\left(\frac{z}{2}\right)- \arctan\left(\frac{z}{x}\right)\right)}\right.\nonumber\\ 
&-\left.(2 m'+\lambda  x+i z) \left(48 \lambda ^4+48m^{\prime 4}+192 \lambda  m^{\prime 3}
+24 m^{\prime 2} \left(12 \lambda ^2-i \lambda  (x-2) z+z^2\right)\right.\right.\nonumber\\
&+\left.\left.12 \lambda
m'\left(16 \lambda ^2-i \lambda  \left(x^2-4\right) z+4 z^2\right)-2 i \lambda  \left(\lambda ^2-1\right) \left(x^3-8\right) y+3 z^4+24 \lambda ^2 z^2\right)\right),\nonumber\\
G_3&=\left(4 \left((2 \lambda +2m'+i z) e^{2 i \lambda  \arctan\left(\frac{z}{2 \Lambda' }\right)}+(-2 \lambda -2 m'+i z) e^{2 i \lambda  \arctan\left(\frac{z}{2}\right)}\right)^2\right)^{-1}\nonumber\\
&\times e^{2 i \lambda  \arctan\left(\frac{z}{2 \Lambda' }\right)} \left(-(\lambda  2\Lambda' -2 m') \left(4 (\lambda +m')^2+z^2\right)\right.\nonumber\\
&\times \left. \exp \left(2 i \lambda  \left(\arctan\left(\frac{z}{x}\right)-\arctan\left(\frac{z}{2\Lambda' }\right)+\arctan\left(\frac{z}{2}\right)\right)\right)\right.\nonumber\\
&+\left.\left(4 (\lambda +m')^2+z^2\right) e^{2 i \lambda \arctan\left(\frac{z}{2}\right)} (-2 \lambda  \Lambda' +6 m'+2 \lambda  x)\right.\nonumber\\
&-\left.(2 m'+\lambda  x) (2 \lambda +2 m'+i z)^2e^{2 i \lambda  \arctan\left(\frac{z}{2\Lambda' }\right)}+(2\lambda  \Lambda' -2 m') (2 \lambda +2m'+i z)^2 e^{2 i \lambda  \arctan\left(\frac{z}{x}\right)}
\nonumber\right.\\
&-\left.(2 \lambda +2m'-i z)^2 (4 m'+\lambda  (x-2 \Lambda ` )) e^{2 i \lambda  \left(2\arctan\left(\frac{z}{2}\right)-\arctan\left(\frac{z}{2\Lambda' }\right)\right)}\right),\nonumber\\
F_3&=-\left(16 \left((z-2i \lambda -2im') e^{2 i \lambda  \arctan\left(\frac{z}{2\Lambda' }\right)}+(z+2i \lambda +2i m') e^{2 i \lambda  \arctan\left(\frac{z}{2}\right)}\right)\right)^{-1}\nonumber\\
&\times \left(-(2 \lambda +2 m'-i z) e^{2 i \lambda  \arctan\left(\frac{z}{2}\right)} \left(12m^{\prime 2}+
m' (-8 \lambda  \Lambda' +8 \lambda  x+4 i z)\right.\right.\nonumber\\
&+\left.\left.\left(\lambda ^2-1\right) x^2-4 \lambda  \Lambda'  (\lambda  x+i z)\right)+(2m'+(\lambda -1) x) (2m'+\lambda  x+x) (2 \lambda +2m'+i z) e^{2 i \lambda 
\arctan\left(\frac{z}{2\Lambda' }\right)}\right.\nonumber\\
&-\left.2 (2\lambda  \Lambda' -2m') (2 \lambda +2m'+i z) (2m'+\lambda  x-i z) e^{2 i \lambda\arctan\left(\frac{z}{x}\right)}\right) .
\end{align}

Knowing the vertex one can calculate the stress tensor two-point function,
diagramaticaly represented as
\begin{center}
\fcolorbox{white}{white}{
  \scalebox{.5}{
  \begin{picture}(576,208) (83,-27)
    \SetWidth{1.0}
    \SetColor{Black}
    \GOval(304,64)(19,19)(0){0.882}
    \GOval(192,128)(16,16)(0){0.882}
    \GOval(192,0)(16,16)(0){0.882}
    \Vertex(96,64){15}
    \Arc[arrow,arrowpos=0.5,arrowlength=5,arrowwidth=2,arrowinset=0.2,clock](178.667,42.667)(85.375,165.53,91.79)
    \Arc[arrow,arrowpos=0.5,arrowlength=5,arrowwidth=2,arrowinset=0.2,clock](178.667,85.333)(85.375,-91.79,-165.53)
    \Arc[arrow,arrowpos=0.5,arrowlength=5,arrowwidth=2,arrowinset=0.2,clock](216,24)(104.307,94.399,32.471)
    \Arc[arrow,arrowpos=0.5,arrowlength=5,arrowwidth=2,arrowinset=0.2,clock](216,104)(104.307,-32.471,-94.399)
    \Text(365,60)[lb]{\Large{\Black{$+$}}}
    \Arc[arrow,arrowpos=0.5,arrowlength=5,arrowwidth=2,arrowinset=0.2,clock](498.667,42.667)(85.375,165.53,91.79)
    \Arc[arrow,arrowpos=0.5,arrowlength=5,arrowwidth=2,arrowinset=0.2,clock](498.667,85.333)(85.375,-91.79,-165.53)
    \GOval(512,128)(16,16)(0){0.882}
    \GOval(512,0)(16,16)(0){0.882}
    \Arc[arrow,arrowpos=0.5,arrowlength=5,arrowwidth=2,arrowinset=0.2,clock](525.333,42.667)(85.375,88.21,14.47)
    \Arc[arrow,arrowpos=0.5,arrowlength=5,arrowwidth=2,arrowinset=0.2,clock](525.333,85.333)(85.375,-14.47,-88.21)
    \Photon(416,64)(608,64){7.5}{10}
    \Text(70,80)[lb]{\Large{\Black{$\lambda \rho$}}}
    \Text(176,160)[lb]{\Large{\Black{$p+q$}}}
    \Text(192,-32)[lb]{\Large{\Black{$p$}}}
    \Text(330,80)[lb]{\Large{\Black{$\mu \nu$}}}
    \Text(512,160)[lb]{\Large{\Black{$p$}}}
    \Text(512,-32)[lb]{\Large{\Black{$k$}}}
    \Text(490,80)[lb]{\Large{\Black{$k+q-p$}}}
    \Text(390,80)[lb]{\Large{\Black{$\lambda \rho$}}}
    \Text(624,80)[lb]{\Large{\Black{$\mu\nu$}}}
  \end{picture}
  }
}
\end{center}
which is equal to
\begin{align}
	\left \langle T^{\mu \nu} (q) T^{\lambda \rho} (-q) \right \rangle  &= \int \frac{d^3 p}{(2 \pi )^3} \text{Tr} \left( G(p+q) U^{\mu \nu} ( p ; q ) G ( p ) U^{\lambda \rho}_0 ( p + q ; - q)  \right) \nonumber \\
	&+ \frac 1 2\int \frac{d^3 p d^3 k}{(2 \pi)^6} G_{\alpha \beta}(k+q-p) \text{Tr} \left( G(p) U^{\mu \nu , \beta}_v G(k) U_v^{\lambda \rho,\alpha}\right) .
\end{align}
Here
\begin{equation}
U^{\mu \nu, \lambda}_v = i (\eta^{\lambda ( \mu} \gamma^{\nu )} - \eta^{\mu \nu} \gamma^\lambda )
\end{equation}
denotes the vertex arising from the $\bar\psi A\psi$ part of the stress tensor. One may check by hand that the final term contributes only to the $+-+-$ component of the correlation function and that that contribution is
\begin{align}
	- \frac 1 4 \int \frac{d^3 p d^3k}{(2 \pi)^6} G_{+3}(k-p)\text{Tr} \left( G(p) H_+ ( G(k) )\right) .
\end{align}

We find the nonzero contributions are
\begin{align}
	\left \langle T^{--} (q) T^{++} (-q) \right \rangle &= \frac{i \mu^3}{8 \pi} \int_2^{2\Lambda'} dx \frac{J_2 + 2 F_2 + X G_2}{x^2+y^2} = - \frac{\mu^3}{8 \pi \lambda} \left( G_2 ( 2\Lambda' , y ) - U_2 ( 2\Lambda' , y ) \right) ,\nonumber \\
	\left \langle T^{+-} (q) T^{+-} (-q) \right \rangle &= \frac{\mu^3}{4 \pi} \int_2^{2 \Lambda'} dx \frac{1}{x^2+y^2} \bigg(\frac{1}{16} \left( Y_1 -2 (x^2 + y^2) \right)^2 - \frac 1 4 \left( 3 x^4 + 4 x^2 y^2 + y^4 \right) \nonumber \\
	& + \left( \frac i 4 X Y_1 - \Sigma'_I (x^2 + y^2 ) \right) U_3 
	- i \left(y^2 + \frac 3 4 x^2 - \Sigma'_I(\Sigma'_I + i y ) \right) V_3 \nonumber \\
	&+\left( - i \left(y^2 + \frac 3 4 x^2 - \Sigma'_I(\Sigma'_I - i y ) \right) + 2 Y U_3 -4 V_3 \right) F_3 \nonumber \\
	&+\left( - \frac i 4 Y Y_1 - \Sigma'_I (x^2 + y^2 ) + 2 Y_1 U_3 -2 X V_3\right) G_3 \bigg) 
	\nonumber \\
	&+ \frac{\lambda \mu^3}{12 \pi} ( \Lambda' - 1)^2( \lambda ( \Lambda' +2) + 3 m' ).
\end{align}

Removing ${\cal O}(\Lambda^n)$, $n=1,2,3$, divergent terms, we obtain
from the $U^{++}$ (or the $U^{--}$) vertex
\begin{align}
\left \langle T^{--} (q) T^{++} (-q) \right \rangle &= \frac{N}{384 \pi  \lambda  \omega }
 \left(4 \lambda  \mu  \omega  \left(12 c_0 (\omega -2 \lambda  \mu )+\left(8 \lambda ^2+4\right) \mu ^2-6 \lambda  \mu  \omega +3 \omega ^2\right)\right.\nonumber\\
 &-\left.3 (\omega -2 c_0) (2 c_0+\omega )^3 \left(1-e^{-2 \lambda  \arctanh \left(\frac{\omega }{2 \mu }\right)}\right)\right)\,.
\end{align}
Here we have Wick rotated back to Lorenztian space to obtain the retarded propagator: $z = - \frac{i \omega}{\mu}$.

Using the $U_{+-}$ vertex we find
\begin{align}
&\left \langle T^{+-} (q) T^{+-} (-q) \right \rangle = 
N\left(48 \pi  \lambda  \left((\omega -2 c_0) e^{2 \lambda  \arctanh \left(\frac{\omega }{2 \mu }\right)}+(2 c_0+\omega )\right)\right)^{-1}\nonumber\\
&\times \left({-}12 c_0^4{-}12 c_0^3 \lambda  \mu {-}3 c_0^2 \left(4 \lambda ^2 \mu ^2{+}6 \lambda  \mu  \omega {-}\omega ^2\right){+}\left(12 c_0^4{+}12 c_0^3 \lambda  \mu {+}3 c_0^2 \left(4 \lambda ^2 \mu ^2{-}6 \lambda  \mu  \omega {-}\omega ^2\right)\nonumber\right.\right.\\
&-\left.\left.2 c_0 \lambda  \mu  \left(2 \left(\lambda ^2+4\right) \mu ^2-3 \omega ^2\right)+\lambda  \mu ^2 \omega  \left(2 \left(\lambda ^2+4\right) \mu -3 \lambda  \omega \right)\right) e^{2 \lambda  \arctanh \left(\frac{\omega }{2 \mu }\right)}\right.\\
&+\left.2 c_0 \lambda  \mu  \left(2 \left(\lambda ^2+4\right) \mu ^2-3 \omega ^2\right)+\lambda  \mu ^2 \omega  \left(2 \left(\lambda ^2+4\right) \mu +3 \lambda  \omega \right)\right)
-\frac{\lambda  \mu ^2 N (\lambda  \mu -3 c_0)}{12 \pi }\,.\nonumber
\end{align}
Regularizing the $T^{+-}$ correlation function is subtle. We have adopted the following
regularization scheme. First the exponents $e^{2\lambda\arctanh\left(\frac{\omega}{\Lambda}\right)}$
were set to one, and then the polynomially divergent terms were removed.

\subsection{Hall Viscosity Contact Terms}
To complete the calculation of the viscosity tensor, we need the contact terms in (\ref{GijklFormula})
\begin{align}\label{TTContact}
	G_\text{contact}^{ij, kl} ( q ) = \left \langle \frac{\delta T^{ij}}{\delta g_{kl} } \right \rangle (q)  .
\end{align}
We begin by considering variations of the spin connection, which will contribute a constant offset to the Hall viscosity. Recall that to couple a Dirac spinor to curved space one needs to introduce a veilbein $e^a_\mu$, that is, a local orthonormal basis for the tangent space.
\begin{align}
	e^a_\mu e^{b \mu} = \eta^{ab},
	&&\text{where}
	&&\eta^{ab} = \text{diag} (-1 ,1, 1 ).
\end{align}
 There are many possible selections of such a basis, related by local Lorentz transformations (LLTs)
\begin{align}
	e^a \rightarrow \left( e^{- \frac i 2 \theta_{cd} J^{cd} } \right)^a{}_b e^b ,
\end{align}
where 
\begin{align}
	(J^{ab})_{cd} = i ( \delta^a{}_d \delta^{b}_{c} - \delta^{a}{}_c \delta^b{}_d ) 
\end{align}
are the generators of $so(2,1)$ in the vector representation. We shall raise and lower Lorentz indices $a,b,\dots$ with the $\eta^{ab}$ and it's inverse $\eta_{ab}$ throughout this section.

Under an LLT, $\psi$ transforms in the Dirac representation
\begin{align}
	\psi \rightarrow e^{- \frac i 2 \theta_{ab} S^{ab} }\psi,
	&&S^{ab} = - \frac i 4 [ \gamma^a , \gamma^b ],
\end{align}
and so the Dirac action involves an $so(d,1)$-valued connection $\omega_\mu^{ab}$ through the Lorentz covariant derivative\footnote{We suppress $A_\mu$ as it is not important here.}
\begin{align}
	D_\mu = \partial_\mu + \frac i 2 \omega_{\mu ab} S^{ab} .
\end{align}
On a metric compatible, torsion free background, this is determined by the veilbein
\begin{align}
	\omega_\mu^{ab} = e^{a \nu} \nabla_\mu e^b_\nu,
\end{align}
from which one may check that $D_\mu \psi$ transforms covariantly under LLTs.

The spin connection then enters the stress through through\footnote{Variation of the spin connection in the action does not contribute extra terms to the stress tensor itself.}
\begin{align}
	T^{\mu \nu} \big|_{\omega \text{ part}} &= \frac i 4 \left( \eta^{\mu \nu} \bar \psi \{ \gamma^\lambda, S^{ab} \} \psi - \eta^{\lambda ( \mu} \bar \psi \{ \gamma^{\nu)} , S^{ab} \} \psi \right) \omega_{\lambda ab} .
\end{align}
Now under a metric variation we choose a gauge where
\begin{align}
	\delta e^\mu_a = - \frac 1 2 \eta^{\mu \lambda} e^\rho_a \delta g_{\lambda \rho},
\end{align}
and an explicit computation gives
\begin{align}
	G^{\mu \nu , \lambda \rho}_\text{contact} (q) &= - \frac i 4 \eta^{( \lambda | ( \mu} \varepsilon^{\nu ) | \rho ) \sigma} q_\sigma \left \langle \bar \psi \psi \right \rangle   .
\end{align}
Rotating back to Minkowski space we have 
\begin{align}
	G^{\mu \nu , \lambda \rho}_\text{contact} (\omega ) &=  \frac i 4 \eta^{( \lambda | ( \mu} \varepsilon^{\nu ) | \rho )} \omega \left \langle \bar \psi \psi \right \rangle ,
\end{align}
where we've defined $\varepsilon^{\mu \nu} = \varepsilon^{\mu \nu 0}$.
The only contribution is to the Hall viscosity
\begin{align}
	&\eta_{H \text{contact}} (\omega ) = \frac{1}{4} \left \langle \bar \psi \psi \right \rangle .
\end{align}
Finally we evaluate this expectation value
\begin{align}
	\left \langle \bar \psi (x) \psi (x) \right \rangle &= - \int \frac{d^3 k}{(2 \pi)^3} \text{Tr}~ G(k) 
	= - \frac{\mu^2}{4 \pi} \int_2^\infty dx \Sigma_I (x).
\end{align}
This is quadratically divergent. Upon regularization we have
\begin{align}
	\left \langle \bar \psi \psi \right \rangle = \frac{\mu}{4 \pi} \left(2 c_0 - \lambda \mu\right).
\end{align}

\subsection{Bulk and Shear Contact Terms}
Now we consider the remaining contributions to the contact term (\ref{TTContact}).
The variation of the stress operator is
\begin{align}
\label{contact}
	\frac{\delta T^{\mu \nu} (x) }{\delta g_{\lambda \rho } (0)} &= \bigg( \frac{1}{2} \bar \psi \gamma^{(\mu} \eta^{\nu ) ( \lambda} \overset{\leftrightarrow} D {}^{\rho)} \psi + \frac{1}{4} \bar \psi \gamma^{(\lambda} \eta^{\rho ) ( \mu} \overset{\leftrightarrow} D {}^{\nu)} \psi \nonumber \\
&- \frac 1 4 \eta^{\mu \nu} \bar \psi \gamma^{(\lambda}\overset{\leftrightarrow} D {}^{\rho)} \psi- \left(\frac 1 2 \bar \psi \gamma^\sigma  \overset{\leftrightarrow} D {}_{\sigma} \psi + m \bar \psi \psi \right) \eta^{\mu ( \lambda} \eta^{\rho) \nu}\bigg) \delta^3 (x).
\end{align}
The contribution to the shear viscosity is easy to evaluate. The $++--$ component of the above is
\begin{align}
	- \frac 1 4 \bar \psi \gamma^- \overset{\leftrightarrow} \partial_- \psi - m \bar \psi \psi,
\end{align}
where we have used the gauge condition $A_-=0$. Taking the expectation value and regulating divergences,
\begin{align}
	\left \langle \frac{\delta T^{++}}{\delta g_{--}} \right \rangle (\omega) &= \int \frac{d^3 k}{(2 \pi)^3} \text{Tr} \left( \left( m + \frac i 2 k_- \gamma^- \right) G(k) \right) \nonumber \\
	&= - \frac{\mu}{24 \pi} \left( 9 c_0^2 - 15 \lambda c_0 \mu + (1+5 \lambda^2) \mu^2 \right) .
\end{align}

Now we turn to the bulk viscosity. The $+-+-$ component of (\ref{contact}) is
\begin{align}
	- \frac 3 {16} \bar \psi \gamma^i \overset{\leftrightarrow} D_i \psi - \frac 1 4 \bar \psi \gamma^3 \overset{\leftrightarrow} D_3 \psi - \frac 1 2 m \bar \psi \psi .
\end{align}
Computing the expectation value proceeds as above but also involves diagrams with a single gauge field.
In the end one finds.
\begin{align}
	&
	G_\text{contact}^{+-, +-}  = 
	\frac{\mu}{48 \pi}\left( (1- \lambda^2 ) \mu^2 +3  \lambda c_0 \mu -3 c_0^2 \right)\,.\nonumber \\
\end{align}

\bibliographystyle{JHEP}
\bibliography{MGbib-v2}

\end{document}